\documentstyle[12pt]{article} 

\begin{document} 

\vspace{15mm}  

\centerline{\Large \bf Integrability Condition in the Statistical Model }
\centerline{\Large \bf  and the Addition Formula of $g=2$ Hyperelliptic Function}

\vspace{10mm}

\centerline{\large Kazuyasu Shigemoto\footnote{E-mail address:
shigemot@tezukayama-u.ac.jp}}

\centerline {{\large
Tezukayama University, Tezukayama 7, Nara 631, Japan
}}

\vspace{10mm}

\centerline{\bf Abstract}  
\noindent 
The integrability condition of the Ising model is understood 
as the $SU(2)$ integrability condition and also as the model
parameterized by the elliptic function, where the integrability
condition is understood as the addition theorem of the elliptic
function.
The generalization of this integrability condition is 
to find the solution of the higher rank Lie group integrability
condition and also to find the model parameterized by higher genus
hyperelliptic function.
For the preparation of this purpose, we give the explicit formula
of the addition formula for $g=2$ hyperelliptic function.

\vspace{5mm}

\noindent
\begin{tabular}{|llll|} 
\hline 
Keywords: & Ising model  , & integrability condition , & theta function\\
   & hyperelliptic function, & addition formula, & Lie group \\
\hline
\end{tabular} 

\vspace{10mm}
\noindent {\bf Contents}\\
I Introduction\\
II Addition Relation of $g=2$ Theta Function\\
III Addition Formula of $g=2$ Hyperelliptic Function\\
IV Summary and Discussion\\
Appendix A: Proof of the Fundamental Addition Relation of $g=2$ Theta Function\\

\newpage

\setcounter{equation}{0}
\section{Introduction}

There are many two dimensional integrable statistical models \cite{Baxter}.
The fundamental integrable statistical model is the Ising model \cite{Onsager}, 
which is parameterized by the elliptic function to satisfy the integrable 
condition.
The general Yang-Baxter equation, which has the difference property, 
for the spin model is given in the form  
\begin{eqnarray}
  && U(x) V(x+y) U(y)=V(y)U(x+y) V(x) .
\label{e1-1}
\end{eqnarray}   
This type of Yang-Baxter equation is called the integrability condition, 
because the Yang-Baxter equation of this type says that the product of three 
group actions for two different paths gives the same group action, 
that is, the product of three group action is path independent. 
In general, if we start from some fixed point in the group
and multiplying the group action through certain path to the final point 
in the group, the total group action is path independent in the above 
integrable model.\\

This integrable model seems to have the connection with the algebraic 
function. The algebraic function has the addition formula. The addition
formula of the trigonometry function can be written as the $U(1)$ integrable
condition in the form Eq.(\ref{e1-1}) through the Euler's relation 
$U(x)=\exp(ix)={\rm cos}(x)+i{\rm sin}(x)$, $V(x)=1$.
The addition formula of the elliptic function can be written as the  
$SU(2)$ Lie group integrable condition in the form Eq.(\ref{e1-1})
\cite{Shigemoto1, Shigemoto2, Shigemoto3} with
\begin{eqnarray}
&&U(x)=\exp\{i{\rm am}(x,k) J_z\},\ V(x)=\exp\{i{\rm am}(kx,1/k) J_x\} , 
\label{e1-2} 
\end{eqnarray}
where $J_x$, $J_z$ are Lie group elements of $SO(3) \cong SU(2)/Z_2$. 
We must notice that we can write the integrable condition in the Lie 
group form by using the elliptic function with some moduli and
the elliptic function with the dual moduli, that is, the Lie group structure 
and the discrete moduli transformation is mutually connected.
\\
There are two directions to generalize the above integrable model.
One direction of the generalization is to consider the 
two variable integrable condition with difference 
property in the form
\begin{eqnarray}
  && U(x_1,y_1) V(x_1+x_2,y_1+y_2) U(x_2,y_2)
    =V(x_2,y_2)U(x_1+x_2,y_1+y_2) V(x_1,y_1) .
\label{e1-3}
\end{eqnarray}
Another direction of the generalization is to consider the higher
genus algebraic function, $g=2$ hyperelliptic function, 
which has two arguments.\\
Then it is quite desirable that the addition formula of 
$g=2$ hyperelliptic function can be written in  the integrable
condition in the from Eq(\ref{e1-3}).\\
If this expectation is true, the integrable condition (path independence) 
of the algebraic function on the Riemann surface,
which is nothing but the addition formula of the hyperelliptic function, 
is expected to be written as the integrable condition of 
certain Lie group.\\
For that preparation, we explicitly derive the addition formula
for $g=2$ hyperelliptic function in this paper.\\
\newpage
\setcounter{equation}{0}
\section{Addition Relation of $g=2$ Theta Function}
The addition relation of $g=2$ hyperelliptic theta function 
is first given by Rosenhain\cite{Rosenhain1, Rosenhain2}, and 
K\"{o}nigsberger\cite{Konigsberger} give the fundamental relation to make the 
addition theorem.\\
The explicit addition formula is given by Kossak\cite{Kossak} but he give only the 
sketch to derive the addition formula and did not give all
the addition formula. Then we will give the detailed derivation 
and give all addition formula.\\
The theta function with two variables is defined by  
\begin{eqnarray}
&&\vartheta\left[\begin{array}{cc} a \ c \\ b \ d \\ 
\end{array}\right](u,v;\tau_1, \tau_2, \tau_{12})
\nonumber\\
&&=\sum_{m,n\in Z} \exp 
\left\{ \pi i  
\Bigl( \tau_1 (m+\frac{a}{2})^2+\tau_2 (n+\frac{c}{2})^2+
2 \tau_{12} (m+\frac{a}{2}) (n+\frac{c}{2}) \Bigr) \right.
\nonumber\\
&& \left. +2\pi i \ 
\Bigl( (m+\frac{a}{2})(u+\frac{b}{2})+(n+\frac{c}{2})(v+\frac{d}{2})
\Bigr)
\right\}  ,
\label{e2-1}
\end{eqnarray}
where we assume that ${\rm Im} \tau_1>0$, ${\rm Im}\tau_2>0$,
$({\rm Im} \tau_1)({\rm Im}\tau_2)-({\rm Im} \tau_{12})^2>0$ in order
that the summation of $m,n \in Z$ becomes convergent. 
We can rename $m\rightarrow m$, $n\rightarrow -n$, so that we can always choose
${\rm Im} \tau_{12}>0$ so we assume ${\rm Im} \tau_{12}>0$.\\
The Riemann's theta relation in this case is given by
\begin{eqnarray}
&&\hskip -13mm\prod_{i=1}^{4} \vartheta\left[
\begin{array}{cc} 0 \ 0 \\ 0 \ 0 \\
\end{array}\right](u_i,v_i)
+\prod_{i=1}^{4} \vartheta\left[\begin{array}{cc} 0 \ 1 \\ 0 \ 0 \\
\end{array}\right](u_i,v_i)
+\prod_{i=1}^{4} \vartheta\left[\begin{array}{cc} 1 \ 0 \\ 0 \ 0 \\
\end{array}\right](u_i,v_i)
+\prod_{i=1}^{4} \vartheta\left[\begin{array}{cc} 1 \ 1 \\ 0 \ 0 \\
\end{array}\right](u_i,v_i)
\nonumber\\
&&\hskip -18mm =\prod_{i=1}^{4} \vartheta\left[\begin{array}{cc} 0 \ 0 \\ 0 \ 0 \\
\end{array}\right](\tilde{u}_i,\tilde{v}_i)
+\prod_{i=1}^{4} \vartheta\left[\begin{array}{cc} 0 \ 1 \\ 0 \ 0 \\
\end{array}\right](\tilde{u}_i,\tilde{v}_i)
+\prod_{i=1}^{4} \vartheta\left[\begin{array}{cc} 1 \ 0 \\ 0 \ 0 \\
\end{array}\right](\tilde{u}_i,\tilde{v}_i)
+\prod_{i=1}^{4} \vartheta\left[\begin{array}{cc} 1 \ 1 \\ 0 \ 0 \\
\end{array}\right](\tilde{u}_i,\tilde{v}_i)  ,
\label{e2-2}\\
&&\hskip -13mm{\rm where}
\nonumber\\
&&\hskip -13mm
\left(\begin{array}{c}
\tilde{u}_1\\
\tilde{u}_2\\
\tilde{u}_3\\
\tilde{u}_4\\
\end{array}\right)
=\left(\begin{array}{c}
(u_1+u_2+u_3+u_4)/2\\
(u_1+u_2-u_3-u_4)/2\\
(u_1-u_2+u_3-u_4)/2\\
(u_1-u_2-u_3+u_4)/2\\
\end{array}\right)  ,
\left(\begin{array}{c}
\tilde{v}_1\\
\tilde{v}_2\\
\tilde{v}_3\\
\tilde{v}_4\\
\end{array}\right)
=\left(\begin{array}{c}
(v_1+v_2+v_3+v_4)/2\\
(v_1+v_2-v_3-v_4)/2\\
(v_1-v_2+v_3-v_4)/2\\
(v_1-v_2-v_3+v_4)/2\\
\end{array}\right) .
\label{e2-3}
\end{eqnarray}
We parameterize in the following way\\
\begin{eqnarray}
&&\hskip -15mm  \vartheta\left[\begin{array}{cc} 0 \ 0 \\ 0 \ 0 \\ 
\end{array}\right](\alpha,\beta)\ 
\vartheta\left[\begin{array}{cc} 0 \ 0 \\ 0 \ 0 \\ 
\end{array}\right](\alpha',\beta')\ 
\vartheta\left[\begin{array}{cc} 0 \ 0 \\ 0 \ 0 \\ 
\end{array}\right](y+y'+\alpha,z+z'+\beta)\ 
\vartheta\left[\begin{array}{cc} 0 \ 0 \\ 0 \ 0 \\ 
\end{array}\right](y-y'+\alpha',z-z'+\beta')
\nonumber\\
&&\hskip -15mm +\vartheta\left[\begin{array}{cc} 0 \ 1 \\ 0 \ 0 \\
\end{array}\right](\alpha,\beta)\ 
\vartheta\left[\begin{array}{cc} 0 \ 1 \\ 0 \ 0 \\
\end{array}\right](\alpha',\beta')\ 
\vartheta\left[\begin{array}{cc} 0 \ 1 \\ 0 \ 0 \\
\end{array}\right](y+y'+\alpha,z+z'+\beta)\ 
\vartheta\left[\begin{array}{cc} 0 \ 1 \\ 0 \ 0 \\
\end{array}\right](y-y'+\alpha',z-z'+\beta')
\nonumber\\
&&\hskip -15mm +\vartheta\left[\begin{array}{cc} 1 \ 0 \\ 0 \ 0 \\
\end{array}\right](\alpha,\beta)\ 
\vartheta\left[\begin{array}{cc} 1 \ 0 \\ 0 \ 0 \\
\end{array}\right](\alpha',\beta')\ 
\vartheta\left[\begin{array}{cc} 1 \ 0 \\ 0 \ 0 \\
\end{array}\right](y+y'+\alpha,z+z'+\beta)\ 
\vartheta\left[\begin{array}{cc} 1 \ 0 \\ 0 \ 0 \\
\end{array}\right](y-y'+\alpha',z-z'+\beta')
\nonumber\\
&&\hskip -15mm +\vartheta\left[\begin{array}{cc} 1 \ 1 \\ 0 \ 0 \\
\end{array}\right](\alpha,\beta)\ 
\vartheta\left[\begin{array}{cc} 1 \ 1 \\ 0 \ 0 \\
\end{array}\right](\alpha',\beta')\ 
\vartheta\left[\begin{array}{cc} 1 \ 1 \\ 0 \ 0 \\
\end{array}\right](y+y'+\alpha,z+z'+\beta)\ 
\vartheta\left[\begin{array}{cc} 1 \ 1 \\ 0 \ 0 \\
\end{array}\right](y-y'+\alpha',z-z'+\beta')
\nonumber\\
&&\hskip -15mm =\vartheta\left[\begin{array}{cc} 0 \ 0 \\ 0 \ 0 \\ 
\end{array}\right](y+\alpha+\alpha',z+\beta+\beta')\ 
\vartheta\left[\begin{array}{cc} 0 \ 0 \\ 0 \ 0 \\ 
\end{array}\right](y,z)\ 
\vartheta\left[\begin{array}{cc} 0 \ 0 \\ 0 \ 0 \\ 
\end{array}\right](y'+\alpha-\alpha',z'+\beta-\beta')\ 
\vartheta\left[\begin{array}{cc} 0 \ 0 \\ 0 \ 0 \\ 
\end{array}\right](y',z')
\nonumber\\
&&\hskip -15mm +\vartheta\left[\begin{array}{cc} 0 \ 1 \\ 0 \ 0 \\
\end{array}\right](y+\alpha+\alpha',z+\beta+\beta')\ 
\vartheta\left[\begin{array}{cc} 0 \ 1 \\ 0 \ 0 \\
\end{array}\right](y,z)\  
\vartheta\left[\begin{array}{cc} 0 \ 1 \\ 0 \ 0 \\
\end{array}\right](y'+\alpha-\alpha',z'+\beta-\beta')\  
\vartheta\left[\begin{array}{cc} 0 \ 1 \\ 0 \ 0 \\
\end{array}\right](y',z')
\nonumber\\
&&\hskip -15mm +\vartheta\left[\begin{array}{cc} 1 \ 0 \\ 0 \ 0 \\
\end{array}\right](y+\alpha+\alpha',z+\beta+\beta')\  
\vartheta\left[\begin{array}{cc} 1 \ 0 \\ 0 \ 0 \\
\end{array}\right](y,z)\  
\vartheta\left[\begin{array}{cc} 1 \ 0 \\ 0 \ 0 \\
\end{array}\right](y'+\alpha-\alpha',z'+\beta-\beta')\  
\vartheta\left[\begin{array}{cc} 1 \ 0 \\ 0 \ 0 \\
\end{array}\right](y',z')
\nonumber\\
&&\hskip -15mm +\vartheta\left[\begin{array}{cc} 1 \ 1 \\ 0 \ 0 \\
\end{array}\right](y+\alpha+\alpha',z+\beta+\beta')\ 
\vartheta\left[\begin{array}{cc} 1 \ 1 \\ 0 \ 0 \\
\end{array}\right](y,z)\ 
\vartheta\left[\begin{array}{cc} 1 \ 1 \\ 0 \ 0 \\
\end{array}\right](y'+\alpha-\alpha',z'+\beta-\beta')\ 
\vartheta\left[\begin{array}{cc} 1 \ 1 \\ 0 \ 0 \\
\end{array}\right](y',z')  . 
\nonumber\\
&&\label{e2-4}
\end{eqnarray}
\noindent Using this relation, we can show(Appendix A) three fundamental
relations in the form.\\
\noindent \underline{Kossak's Fundamental Relation 1):}\\ 
\begin{eqnarray}
&&\hskip -15mm  \vartheta\left[\begin{array}{cc} 0 \ 0 \\ 0 \ 0 \\ 
\end{array}\right](\alpha,\beta)\ 
\vartheta\left[\begin{array}{cc} 0 \ 0 \\ 1 \ 1 \\ 
\end{array}\right](0,0)\ 
\vartheta\left[\begin{array}{cc} 0 \ 0 \\ 0 \ 0 \\ 
\end{array}\right](y+y'+\alpha,z+z'+\beta)\ 
\vartheta\left[\begin{array}{cc} 0 \ 0 \\ 1 \ 1 \\ 
\end{array}\right](y-y',z-z')
\nonumber\\
&&\hskip -15mm =\vartheta\left[\begin{array}{cc} 0 \ 0 \\ 0 \ 0 \\
\end{array}\right](y+\alpha,z+\beta)\ 
\vartheta\left[\begin{array}{cc} 0 \ 0 \\ 1 \ 1 \\
\end{array}\right](y,z)\ 
\vartheta\left[\begin{array}{cc} 0 \ 0 \\ 0 \ 0 \\
\end{array}\right](y'+\alpha,z'+\beta)\ 
\vartheta\left[\begin{array}{cc} 0 \ 0 \\ 1 \ 1 \\
\end{array}\right](y',z')
\nonumber\\
&&\hskip -15mm -\vartheta\left[\begin{array}{cc} 1 \ 0 \\ 0 \ 0 \\
\end{array}\right](y+\alpha,z+\beta)\ 
\vartheta\left[\begin{array}{cc} 1 \ 0 \\ 1 \ 1 \\
\end{array}\right](y,z)\ 
\vartheta\left[\begin{array}{cc} 1 \ 0 \\ 0 \ 0 \\
\end{array}\right](y'+\alpha,z'+\beta)\ 
\vartheta\left[\begin{array}{cc} 1 \ 0 \\ 1 \ 1 \\
\end{array}\right](y',z')
\nonumber\\
&&\hskip -15mm -\vartheta\left[\begin{array}{cc} 0 \ 1 \\ 1 \ 0 \\
\end{array}\right](y+\alpha,z+\beta)\ 
\vartheta\left[\begin{array}{cc} 0 \ 1 \\ 0 \ 1 \\
\end{array}\right](y,z)\ 
\vartheta\left[\begin{array}{cc} 0 \ 1 \\ 1 \ 0 \\
\end{array}\right](y'+\alpha,z'+\beta)\ 
\vartheta\left[\begin{array}{cc} 0 \ 1 \\ 0 \ 1 \\
\end{array}\right](y',z')
\nonumber\\
&&\hskip -15mm +\vartheta\left[\begin{array}{cc} 1 \ 1 \\ 1 \ 0 \\ 
\end{array}\right](y+\alpha,z+\beta)\ 
\vartheta\left[\begin{array}{cc} 1 \ 1 \\ 0 \ 1 \\ 
\end{array}\right](y,z)\ 
\vartheta\left[\begin{array}{cc} 1 \ 1 \\ 1 \ 0 \\ 
\end{array}\right](y'+\alpha,z'+\beta)\ 
\vartheta\left[\begin{array}{cc} 1 \ 1 \\ 0 \ 1 \\ 
\end{array}\right](y',z') . 
\label{e2-5}
\end{eqnarray}
\noindent \underline{Kossak's Fundamental Relation 2):}\\ 
\begin{eqnarray}
&&\hskip -15mm  \vartheta\left[\begin{array}{cc} 0 \ 0 \\ 1 \ 0 \\ 
\end{array}\right](\alpha,\beta)\ 
\vartheta\left[\begin{array}{cc} 0 \ 0 \\ 0 \ 1 \\ 
\end{array}\right](0,0)\ 
\vartheta\left[\begin{array}{cc} 0 \ 0 \\ 0 \ 0 \\ 
\end{array}\right](y+y'+\alpha,z+z'+\beta)\ 
\vartheta\left[\begin{array}{cc} 0 \ 0 \\ 1 \ 1 \\ 
\end{array}\right](y-y',z-z')
\nonumber\\
&&\hskip -15mm =\vartheta\left[\begin{array}{cc} 0 \ 0 \\ 1 \ 0 \\
\end{array}\right](y+\alpha,z+\beta)\ 
\vartheta\left[\begin{array}{cc} 0 \ 0 \\ 0 \ 1 \\
\end{array}\right](y,z)\ 
\vartheta\left[\begin{array}{cc} 0 \ 0 \\ 0 \ 0 \\
\end{array}\right](y'+\alpha,z'+\beta)\ 
\vartheta\left[\begin{array}{cc} 0 \ 0 \\ 1 \ 1 \\
\end{array}\right](y',z')
\nonumber\\
&&\hskip -15mm -\vartheta\left[\begin{array}{cc} 0 \ 1 \\ 1 \ 0 \\
\end{array}\right](y+\alpha,z+\beta)\ 
\vartheta\left[\begin{array}{cc} 0 \ 1 \\ 0 \ 1 \\
\end{array}\right](y,z)\ 
\vartheta\left[\begin{array}{cc} 0 \ 1 \\ 0 \ 0 \\
\end{array}\right](y'+\alpha,z'+\beta)\ 
\vartheta\left[\begin{array}{cc} 0 \ 1 \\ 1 \ 1 \\
\end{array}\right](y',z')
\nonumber\\
&&\hskip -15mm +\vartheta\left[\begin{array}{cc} 0 \ 0 \\ 0 \ 0 \\
\end{array}\right](y+\alpha,z+\beta)\ 
\vartheta\left[\begin{array}{cc} 0 \ 0 \\ 1 \ 1 \\
\end{array}\right](y,z)\ 
\vartheta\left[\begin{array}{cc} 0 \ 0 \\ 1 \ 0 \\
\end{array}\right](y'+\alpha,z'+\beta)\ 
\vartheta\left[\begin{array}{cc} 0 \ 0 \\ 0 \ 1 \\
\end{array}\right](y',z')
\nonumber\\
&&\hskip -15mm -\vartheta\left[\begin{array}{cc} 0 \ 1 \\ 0 \ 0 \\ 
\end{array}\right](y+\alpha,z+\beta)\ 
\vartheta\left[\begin{array}{cc} 0 \ 1 \\ 1 \ 1 \\ 
\end{array}\right](y,z)\ 
\vartheta\left[\begin{array}{cc} 0 \ 1 \\ 1 \ 0 \\ 
\end{array}\right](y'+\alpha,z'+\beta)\ 
\vartheta\left[\begin{array}{cc} 0 \ 1 \\ 0 \ 1 \\ 
\end{array}\right](y',z') ,  
\label{e2-6}\\
&&\hskip -15mm {\rm for\ such}\ \alpha, \beta\ {\rm as\ } 
\vartheta\left[\begin{array}{cc} 0 \ 0 \\ 0 \ 0 \\
\end{array}\right](\alpha,\beta)=0.
\nonumber
\end{eqnarray}
\noindent \underline{Kossak's Fundamental Relation 3):}\\ 
\begin{eqnarray}
&&\hskip -15mm  \vartheta\left[\begin{array}{cc} 0 \ 0 \\ 0 \ 1 \\ 
\end{array}\right](\alpha,\beta)\ 
\vartheta\left[\begin{array}{cc} 0 \ 0 \\ 1 \ 0 \\ 
\end{array}\right](0,0)\ 
\vartheta\left[\begin{array}{cc} 0 \ 0 \\ 0 \ 0 \\ 
\end{array}\right](y+y'+\alpha,z+z'+\beta)\ 
\vartheta\left[\begin{array}{cc} 0 \ 0 \\ 1 \ 1 \\ 
\end{array}\right](y-y',z-z')
\nonumber\\
&&\hskip -15mm =\vartheta\left[\begin{array}{cc} 0 \ 0 \\ 0 \ 1 \\
\end{array}\right](y+\alpha,z+\beta)\ 
\vartheta\left[\begin{array}{cc} 0 \ 0 \\ 1 \ 0 \\
\end{array}\right](y,z)\ 
\vartheta\left[\begin{array}{cc} 0 \ 0 \\ 0 \ 0 \\
\end{array}\right](y'+\alpha,z'+\beta)\ 
\vartheta\left[\begin{array}{cc} 0 \ 0 \\ 1 \ 1 \\
\end{array}\right](y',z')
\nonumber\\
&&\hskip -15mm -\vartheta\left[\begin{array}{cc} 1 \ 0 \\ 0 \ 1 \\
\end{array}\right](y+\alpha,z+\beta)\ 
\vartheta\left[\begin{array}{cc} 1 \ 0 \\ 1 \ 0 \\
\end{array}\right](y,z)\ 
\vartheta\left[\begin{array}{cc} 1 \ 0 \\ 0 \ 0 \\
\end{array}\right](y'+\alpha,z'+\beta)\ 
\vartheta\left[\begin{array}{cc} 1 \ 0 \\ 1 \ 1 \\
\end{array}\right](y',z')
\nonumber\\
&&\hskip -15mm +\vartheta\left[\begin{array}{cc} 0 \ 0 \\ 0 \ 0 \\
\end{array}\right](y+\alpha,z+\beta)\ 
\vartheta\left[\begin{array}{cc} 0 \ 0 \\ 1 \ 1 \\
\end{array}\right](y,z)\ 
\vartheta\left[\begin{array}{cc} 0 \ 0 \\ 0 \ 1 \\
\end{array}\right](y'+\alpha,z'+\beta)\ 
\vartheta\left[\begin{array}{cc} 0 \ 0 \\ 1 \ 0 \\
\end{array}\right](y',z')
\nonumber\\
&&\hskip -15mm -\vartheta\left[\begin{array}{cc} 1 \ 0 \\ 0 \ 0 \\ 
\end{array}\right](y+\alpha,z+\beta)\ 
\vartheta\left[\begin{array}{cc} 1 \ 0 \\ 1 \ 1 \\ 
\end{array}\right](y,z)\ 
\vartheta\left[\begin{array}{cc} 1 \ 0 \\ 0 \ 1 \\ 
\end{array}\right](y'+\alpha,z'+\beta)\ 
\vartheta\left[\begin{array}{cc} 1 \ 0 \\ 1 \ 0 \\ 
\end{array}\right](y',z')
\label{e2-7}\\
&&\hskip -15mm {\rm for\ such\ }\ \alpha, \beta\ {\rm as\ } 
\vartheta\left[\begin{array}{cc} 0 \ 0 \\ 0 \ 0 \\
\end{array}\right](\alpha,\beta)=0 . 
\nonumber
\end{eqnarray}
\noindent \underline{Addition Formula 1):}\\ 
\noindent Putting $\alpha=1/2$, $\beta=1/2$ in Eq.(\ref{e2-5}), we have 
\begin{eqnarray}
&&\hskip -15mm  \vartheta^2 \left[\begin{array}{cc} 0 \ 0 \\ 1 \ 1 \\ 
\end{array}\right](0,0)\ 
\vartheta\left[\begin{array}{cc} 0 \ 0 \\ 1 \ 1 \\ 
\end{array}\right](y+y',z+z')\ 
\vartheta\left[\begin{array}{cc} 0 \ 0 \\ 1 \ 1 \\ 
\end{array}\right](y-y',z-z')
\nonumber\\
&&\hskip -15mm 
=\vartheta^2 \left[\begin{array}{cc} 0 \ 0 \\ 1 \ 1 \\
\end{array}\right](y,z)\ 
\vartheta^2 \left[\begin{array}{cc} 0 \ 0 \\ 1 \ 1 \\
\end{array}\right](y',z')\ 
-\vartheta^2 \left[\begin{array}{cc} 1 \ 0 \\ 1 \ 1 \\
\end{array}\right](y,z)\ 
\vartheta^2 \left[\begin{array}{cc} 1 \ 0 \\ 1 \ 1 \\
\end{array}\right](y',z')
\nonumber\\
&&\hskip -15mm 
-\vartheta^2 \left[\begin{array}{cc} 0 \ 1 \\ 0 \ 1 \\
\end{array}\right](y,z)\ 
\vartheta^2\left[\begin{array}{cc} 0 \ 1 \\ 0 \ 1 \\
\end{array}\right](y',z')\ 
+\vartheta^2 \left[\begin{array}{cc} 1 \ 1 \\ 0 \ 1 \\ 
\end{array}\right](y,z)\ 
\vartheta\left[\begin{array}{cc} 1 \ 1 \\ 0 \ 1 \\ 
\end{array}\right](y',z') . 
\label{e2-8}
\end{eqnarray}
\noindent \underline{Addition Formula 2):}\\ 
\noindent Putting $\alpha=0$, $\beta=1/2$ in Eq.(\ref{e2-5}), we have
\begin{eqnarray}
&&\hskip -15mm  
\vartheta\left[\begin{array}{cc} 0 \ 0 \\ 0 \ 1 \\ 
\end{array}\right](0,0)\ 
\vartheta\left[\begin{array}{cc} 0 \ 0 \\ 1 \ 1 \\ 
\end{array}\right](0,0)\
\vartheta\left[\begin{array}{cc} 0 \ 0 \\ 0 \ 1 \\ 
\end{array}\right](y+y',z+z')\ 
\vartheta\left[\begin{array}{cc} 0 \ 0 \\ 1 \ 1 \\ 
\end{array}\right](y-y',z-z')
\nonumber\\
&&\hskip -15mm 
=\vartheta \left[\begin{array}{cc} 0 \ 0 \\ 1 \ 1 \\
\end{array}\right](y,z)\ 
\vartheta \left[\begin{array}{cc} 0 \ 0 \\ 0 \ 1 \\
\end{array}\right](y,z)\ 
\vartheta \left[\begin{array}{cc} 0 \ 0 \\ 1 \ 1 \\
\end{array}\right](y',z')\ 
\vartheta \left[\begin{array}{cc} 0 \ 0 \\ 0 \ 1 \\
\end{array}\right](y',z')\ 
\nonumber\\
&&\hskip -15mm 
-\vartheta \left[\begin{array}{cc} 1 \ 0 \\ 1 \ 1 \\
\end{array}\right](y,z)\ 
\vartheta \left[\begin{array}{cc} 1 \ 0 \\ 0 \ 1 \\
\end{array}\right](y,z)\ 
\vartheta \left[\begin{array}{cc} 1 \ 0 \\ 1 \ 1 \\
\end{array}\right](y',z')\ 
\vartheta \left[\begin{array}{cc} 1 \ 0 \\ 0 \ 1 \\
\end{array}\right](y',z')\ 
\nonumber\\
&&\hskip -15mm 
-\vartheta \left[\begin{array}{cc} 0 \ 1 \\ 0 \ 1 \\
\end{array}\right](y,z)\ 
\vartheta \left[\begin{array}{cc} 0 \ 1 \\ 1 \ 1 \\
\end{array}\right](y,z)\ 
\vartheta \left[\begin{array}{cc} 0 \ 1 \\ 0 \ 1 \\
\end{array}\right](y',z')\ 
\vartheta \left[\begin{array}{cc} 0 \ 1 \\ 1 \ 1 \\
\end{array}\right](y',z')\ 
\nonumber\\
&&\hskip -15mm 
+\vartheta \left[\begin{array}{cc} 1 \ 1 \\ 0 \ 1 \\
\end{array}\right](y,z)\ 
\vartheta \left[\begin{array}{cc} 1 \ 1 \\ 1 \ 1 \\
\end{array}\right](y,z)\ 
\vartheta \left[\begin{array}{cc} 1 \ 1 \\ 0 \ 1 \\
\end{array}\right](y',z')\ 
\vartheta \left[\begin{array}{cc} 1 \ 1 \\ 1 \ 1 \\
\end{array}\right](y',z')\ .  
\label{e2-9}
\end{eqnarray} 
\noindent \underline{Addition Formula 3):}\\ 
\noindent Putting $\alpha=1/2$, $\beta=0$ in Eq.(\ref{e2-5}), we have
\begin{eqnarray}
&&\hskip -15mm  
\vartheta\left[\begin{array}{cc} 0 \ 0 \\ 1 \ 0 \\ 
\end{array}\right](0,0)\ 
\vartheta\left[\begin{array}{cc} 0 \ 0 \\ 1 \ 1 \\ 
\end{array}\right](0,0)\
\vartheta\left[\begin{array}{cc} 0 \ 0 \\ 1 \ 0 \\ 
\end{array}\right](y+y',z+z')\ 
\vartheta\left[\begin{array}{cc} 0 \ 0 \\ 1 \ 1 \\ 
\end{array}\right](y-y',z-z')
\nonumber\\
&&\hskip -15mm 
=\vartheta \left[\begin{array}{cc} 0 \ 0 \\ 1 \ 1 \\
\end{array}\right](y,z)\ 
\vartheta \left[\begin{array}{cc} 0 \ 0 \\ 1 \ 0 \\
\end{array}\right](y,z)\ 
\vartheta \left[\begin{array}{cc} 0 \ 0 \\ 1 \ 1 \\
\end{array}\right](y',z')\ 
\vartheta \left[\begin{array}{cc} 0 \ 0 \\ 1 \ 0 \\
\end{array}\right](y',z')\ 
\nonumber\\
&&\hskip -15mm 
-\vartheta \left[\begin{array}{cc} 1 \ 0 \\ 1 \ 1 \\
\end{array}\right](y,z)\ 
\vartheta \left[\begin{array}{cc} 1 \ 0 \\ 1 \ 0 \\
\end{array}\right](y,z)\ 
\vartheta \left[\begin{array}{cc} 1 \ 0 \\ 1 \ 1 \\
\end{array}\right](y',z')\ 
\vartheta \left[\begin{array}{cc} 1 \ 0 \\ 1 \ 0 \\
\end{array}\right](y',z')\ 
\nonumber\\
&&\hskip -15mm 
-\vartheta \left[\begin{array}{cc} 0 \ 1 \\ 0 \ 1 \\
\end{array}\right](y,z)\ 
\vartheta \left[\begin{array}{cc} 0 \ 1 \\ 0 \ 0 \\
\end{array}\right](y,z)\ 
\vartheta \left[\begin{array}{cc} 0 \ 1 \\ 0 \ 1 \\
\end{array}\right](y',z')\ 
\vartheta \left[\begin{array}{cc} 0 \ 1 \\ 0 \ 0 \\
\end{array}\right](y',z')\ 
\nonumber\\
&&\hskip -15mm 
+\vartheta \left[\begin{array}{cc} 1 \ 1 \\ 0 \ 1 \\
\end{array}\right](y,z)\ 
\vartheta \left[\begin{array}{cc} 1 \ 1 \\ 0 \ 0 \\
\end{array}\right](y,z)\ 
\vartheta \left[\begin{array}{cc} 1 \ 1 \\ 0 \ 1 \\
\end{array}\right](y',z')\ 
\vartheta \left[\begin{array}{cc} 1 \ 1 \\ 0 \ 0 \\
\end{array}\right](y',z')  . 
\label{e2-10}
\end{eqnarray} 
\noindent \underline{Addition Formula 4):}\\ 
\noindent Putting $\alpha=\tau_1/2$, $\beta=\tau_{12}/2+1/2$ in Eq.(\ref{e2-5}), 
we have 
\begin{eqnarray}
&&\hskip -15mm  
\vartheta\left[\begin{array}{cc} 1 \ 0 \\ 0 \ 1 \\ 
\end{array}\right](0,0)\ 
\vartheta\left[\begin{array}{cc} 0 \ 0 \\ 1 \ 1 \\ 
\end{array}\right](0,0)\
\vartheta\left[\begin{array}{cc} 1 \ 0 \\ 0 \ 1 \\ 
\end{array}\right](y+y',z+z')\ 
\vartheta\left[\begin{array}{cc} 0 \ 0 \\ 1 \ 1 \\ 
\end{array}\right](y-y',z-z')
\nonumber\\
&&\hskip -15mm 
=\vartheta \left[\begin{array}{cc} 0 \ 0 \\ 1 \ 1 \\
\end{array}\right](y,z)\ 
\vartheta \left[\begin{array}{cc} 1 \ 0 \\ 0 \ 1 \\
\end{array}\right](y,z)\ 
\vartheta \left[\begin{array}{cc} 0 \ 0 \\ 1 \ 1 \\
\end{array}\right](y',z')\ 
\vartheta \left[\begin{array}{cc} 1 \ 0 \\ 0 \ 1 \\
\end{array}\right](y',z')\ 
\nonumber\\
&&\hskip -15mm 
-\vartheta \left[\begin{array}{cc} 1 \ 0 \\ 1 \ 1 \\
\end{array}\right](y,z)\ 
\vartheta \left[\begin{array}{cc} 0 \ 0 \\ 0 \ 1 \\
\end{array}\right](y,z)\ 
\vartheta \left[\begin{array}{cc} 1 \ 0 \\ 1 \ 1 \\
\end{array}\right](y',z')\ 
\vartheta \left[\begin{array}{cc} 0 \ 0 \\ 0 \ 1 \\
\end{array}\right](y',z')\ 
\nonumber\\
&&\hskip -15mm 
+\vartheta \left[\begin{array}{cc} 0 \ 1 \\ 0 \ 1 \\
\end{array}\right](y,z)\ 
\vartheta \left[\begin{array}{cc} 1 \ 1 \\ 1 \ 1 \\
\end{array}\right](y,z)\ 
\vartheta \left[\begin{array}{cc} 0 \ 1 \\ 0 \ 1 \\
\end{array}\right](y',z')\ 
\vartheta \left[\begin{array}{cc} 1 \ 1 \\ 1 \ 1 \\
\end{array}\right](y',z')\ 
\nonumber\\
&&\hskip -15mm 
-\vartheta \left[\begin{array}{cc} 1 \ 1 \\ 0 \ 1 \\
\end{array}\right](y,z)\ 
\vartheta \left[\begin{array}{cc} 0 \ 1 \\ 1 \ 1 \\
\end{array}\right](y,z)\ 
\vartheta \left[\begin{array}{cc} 1 \ 1 \\ 0 \ 1 \\
\end{array}\right](y',z')\ 
\vartheta \left[\begin{array}{cc} 0 \ 1 \\ 1 \ 1 \\
\end{array}\right](y',z') . 
\label{e2-11}
\end{eqnarray} 
\noindent \underline{Addition Formula 5):}\\ 
\noindent Putting $\alpha=\tau_{12}/2+1/2$, 
$\beta=\tau_2/2$ in Eq.(\ref{e2-5}), 
we have 
\begin{eqnarray}
&&\hskip -15mm  
\vartheta\left[\begin{array}{cc} 0 \ 1 \\ 1 \ 0 \\ 
\end{array}\right](0,0)\ 
\vartheta\left[\begin{array}{cc} 0 \ 0 \\ 1 \ 1 \\ 
\end{array}\right](0,0)\
\vartheta\left[\begin{array}{cc} 0 \ 1 \\ 1 \ 0 \\ 
\end{array}\right](y+y',z+z')\ 
\vartheta\left[\begin{array}{cc} 0 \ 0 \\ 1 \ 1 \\ 
\end{array}\right](y-y',z-z')
\nonumber\\
&&\hskip -15mm 
=\vartheta \left[\begin{array}{cc} 0 \ 0 \\ 1 \ 1 \\
\end{array}\right](y,z)\ 
\vartheta \left[\begin{array}{cc} 0 \ 1 \\ 1 \ 0 \\
\end{array}\right](y,z)\ 
\vartheta \left[\begin{array}{cc} 0 \ 0 \\ 1 \ 1 \\
\end{array}\right](y',z')\ 
\vartheta \left[\begin{array}{cc} 0 \ 1 \\ 1 \ 0 \\
\end{array}\right](y',z')\ 
\nonumber\\
&&\hskip -15mm 
-\vartheta \left[\begin{array}{cc} 1 \ 0 \\ 1 \ 1 \\
\end{array}\right](y,z)\ 
\vartheta \left[\begin{array}{cc} 1 \ 1 \\ 1 \ 0 \\
\end{array}\right](y,z)\ 
\vartheta \left[\begin{array}{cc} 1 \ 0 \\ 1 \ 1 \\
\end{array}\right](y',z')\ 
\vartheta \left[\begin{array}{cc} 1 \ 1 \\ 1 \ 0 \\
\end{array}\right](y',z')\ 
\nonumber\\
&&\hskip -15mm 
-\vartheta \left[\begin{array}{cc} 0 \ 1 \\ 0 \ 1 \\
\end{array}\right](y,z)\ 
\vartheta \left[\begin{array}{cc} 0 \ 0 \\ 0 \ 0 \\
\end{array}\right](y,z)\ 
\vartheta \left[\begin{array}{cc} 0 \ 1 \\ 0 \ 1 \\
\end{array}\right](y',z')\ 
\vartheta \left[\begin{array}{cc} 0 \ 0 \\ 0 \ 0 \\
\end{array}\right](y',z')\ 
\nonumber\\
&&\hskip -15mm 
+\vartheta \left[\begin{array}{cc} 1 \ 1 \\ 0 \ 1 \\
\end{array}\right](y,z)\ 
\vartheta \left[\begin{array}{cc} 1 \ 0 \\ 0 \ 0 \\
\end{array}\right](y,z)\ 
\vartheta \left[\begin{array}{cc} 1 \ 1 \\ 0 \ 1 \\
\end{array}\right](y',z')\ 
\vartheta \left[\begin{array}{cc} 1 \ 0 \\ 0 \ 0 \\
\end{array}\right](y',z') . 
\label{e2-12}
\end{eqnarray} 
\noindent \underline{Addition Formula 6):}\\ 
\noindent Putting $\alpha=\tau_{12}/2$, 
$\beta=\tau_2/2$ in Eq.(\ref{e2-5}), 
we have
\begin{eqnarray}
&&\hskip -15mm  
\vartheta\left[\begin{array}{cc} 0 \ 1 \\ 0 \ 0 \\ 
\end{array}\right](0,0)\ 
\vartheta\left[\begin{array}{cc} 0 \ 0 \\ 1 \ 1 \\ 
\end{array}\right](0,0)\
\vartheta\left[\begin{array}{cc} 0 \ 1 \\ 0 \ 0 \\ 
\end{array}\right](y+y',z+z')\ 
\vartheta\left[\begin{array}{cc} 0 \ 0 \\ 1 \ 1 \\ 
\end{array}\right](y-y',z-z')
\nonumber\\
&&\hskip -15mm 
=\vartheta \left[\begin{array}{cc} 0 \ 0 \\ 1 \ 1 \\
\end{array}\right](y,z)\ 
\vartheta \left[\begin{array}{cc} 0 \ 1 \\ 0 \ 0 \\
\end{array}\right](y,z)\ 
\vartheta \left[\begin{array}{cc} 0 \ 0 \\ 1 \ 1 \\
\end{array}\right](y',z')\ 
\vartheta \left[\begin{array}{cc} 0 \ 1 \\ 0 \ 0 \\
\end{array}\right](y',z')\ 
\nonumber\\
&&\hskip -15mm 
-\vartheta \left[\begin{array}{cc} 1 \ 0 \\ 1 \ 1 \\
\end{array}\right](y,z)\ 
\vartheta \left[\begin{array}{cc} 1 \ 1 \\ 0 \ 0 \\
\end{array}\right](y,z)\ 
\vartheta \left[\begin{array}{cc} 1 \ 0 \\ 1 \ 1 \\
\end{array}\right](y',z')\ 
\vartheta \left[\begin{array}{cc} 1 \ 1 \\ 0 \ 0 \\
\end{array}\right](y',z')\ 
\nonumber\\
&&\hskip -15mm 
-\vartheta \left[\begin{array}{cc} 0 \ 1 \\ 0 \ 1 \\
\end{array}\right](y,z)\ 
\vartheta \left[\begin{array}{cc} 0 \ 0 \\ 1 \ 0 \\
\end{array}\right](y,z)\ 
\vartheta \left[\begin{array}{cc} 0 \ 1 \\ 0 \ 1 \\
\end{array}\right](y',z')\ 
\vartheta \left[\begin{array}{cc} 0 \ 0 \\ 1 \ 0 \\
\end{array}\right](y',z')\ 
\nonumber\\
&&\hskip -15mm 
+\vartheta \left[\begin{array}{cc} 1 \ 1 \\ 0 \ 1 \\
\end{array}\right](y,z)\ 
\vartheta \left[\begin{array}{cc} 1 \ 0 \\ 1 \ 0 \\
\end{array}\right](y,z)\ 
\vartheta \left[\begin{array}{cc} 1 \ 1 \\ 0 \ 1 \\
\end{array}\right](y',z')\ 
\vartheta \left[\begin{array}{cc} 1 \ 0 \\ 1 \ 0 \\
\end{array}\right](y',z')  .  
\label{e2-13}
\end{eqnarray} 
\noindent \underline{Addition Formula 7):}\\ 
\noindent Putting $\alpha=\tau_1/2$, 
$\beta=\tau_{12}/2$ in Eq.(\ref{e2-5}), 
we have
\begin{eqnarray}
&&\hskip -15mm  
\vartheta\left[\begin{array}{cc} 1 \ 0 \\ 0 \ 0 \\ 
\end{array}\right](0,0)\ 
\vartheta\left[\begin{array}{cc} 0 \ 0 \\ 1 \ 1 \\ 
\end{array}\right](0,0)\
\vartheta\left[\begin{array}{cc} 1 \ 0 \\ 0 \ 0 \\ 
\end{array}\right](y+y',z+z')\ 
\vartheta\left[\begin{array}{cc} 0 \ 0 \\ 1 \ 1 \\ 
\end{array}\right](y-y',z-z')
\nonumber\\
&&\hskip -15mm 
=\vartheta \left[\begin{array}{cc} 0 \ 0 \\ 1 \ 1 \\
\end{array}\right](y,z)\ 
\vartheta \left[\begin{array}{cc} 1 \ 0 \\ 0 \ 0 \\
\end{array}\right](y,z)\ 
\vartheta \left[\begin{array}{cc} 0 \ 0 \\ 1 \ 1 \\
\end{array}\right](y',z')\ 
\vartheta \left[\begin{array}{cc} 1 \ 0 \\ 0 \ 0 \\
\end{array}\right](y',z')\ 
\nonumber\\
&&\hskip -15mm 
-\vartheta \left[\begin{array}{cc} 1 \ 0 \\ 1 \ 1 \\
\end{array}\right](y,z)\ 
\vartheta \left[\begin{array}{cc} 0 \ 0 \\ 0 \ 0 \\
\end{array}\right](y,z)\ 
\vartheta \left[\begin{array}{cc} 1 \ 0 \\ 1 \ 1 \\
\end{array}\right](y',z')\ 
\vartheta \left[\begin{array}{cc} 0 \ 0 \\ 0 \ 0 \\
\end{array}\right](y',z')\ 
\nonumber\\
&&\hskip -15mm 
+\vartheta \left[\begin{array}{cc} 0 \ 1 \\ 0 \ 1 \\
\end{array}\right](y,z)\ 
\vartheta \left[\begin{array}{cc} 1 \ 1 \\ 1 \ 0 \\
\end{array}\right](y,z)\ 
\vartheta \left[\begin{array}{cc} 0 \ 1 \\ 0 \ 1 \\
\end{array}\right](y',z')\ 
\vartheta \left[\begin{array}{cc} 1 \ 1 \\ 1 \ 0 \\
\end{array}\right](y',z')\ 
\nonumber\\
&&\hskip -15mm 
-\vartheta \left[\begin{array}{cc} 1 \ 1 \\ 0 \ 1 \\
\end{array}\right](y,z)\ 
\vartheta \left[\begin{array}{cc} 0 \ 1 \\ 1 \ 0 \\
\end{array}\right](y,z)\ 
\vartheta \left[\begin{array}{cc} 1 \ 1 \\ 0 \ 1 \\
\end{array}\right](y',z')\ 
\vartheta \left[\begin{array}{cc} 0 \ 1 \\ 1 \ 0 \\
\end{array}\right](y',z')  . 
\label{e2-14}
\end{eqnarray} 
\noindent \underline{Addition Formula 8):}\\ 
\noindent Putting $\alpha=\tau_1/2+\tau_{12}/2$, 
$\beta=\tau_2/2+\tau_{12}/2$ in Eq.(\ref{e2-5}), 
we have
\begin{eqnarray}
&&\hskip -15mm  
\vartheta\left[\begin{array}{cc} 1 \ 1 \\ 0 \ 0 \\ 
\end{array}\right](0,0)\ 
\vartheta\left[\begin{array}{cc} 0 \ 0 \\ 1 \ 1 \\ 
\end{array}\right](0,0)\
\vartheta\left[\begin{array}{cc} 1 \ 1 \\ 0 \ 0 \\ 
\end{array}\right](y+y',z+z')\ 
\vartheta\left[\begin{array}{cc} 0 \ 0 \\ 1 \ 1 \\ 
\end{array}\right](y-y',z-z')
\nonumber\\
&&\hskip -15mm 
=\vartheta \left[\begin{array}{cc} 0 \ 0 \\ 1 \ 1 \\
\end{array}\right](y,z)\ 
\vartheta \left[\begin{array}{cc} 1 \ 1 \\ 0 \ 0 \\
\end{array}\right](y,z)\ 
\vartheta \left[\begin{array}{cc} 0 \ 0 \\ 1 \ 1 \\
\end{array}\right](y',z')\ 
\vartheta \left[\begin{array}{cc} 1 \ 1 \\ 0 \ 0 \\
\end{array}\right](y',z')\ 
\nonumber\\
&&\hskip -15mm 
-\vartheta \left[\begin{array}{cc} 1 \ 0 \\ 1 \ 1 \\
\end{array}\right](y,z)\ 
\vartheta \left[\begin{array}{cc} 0 \ 1 \\ 0 \ 0 \\
\end{array}\right](y,z)\ 
\vartheta \left[\begin{array}{cc} 1 \ 0 \\ 1 \ 1 \\
\end{array}\right](y',z')\ 
\vartheta \left[\begin{array}{cc} 0 \ 1 \\ 0 \ 0 \\
\end{array}\right](y',z')\ 
\nonumber\\
&&\hskip -15mm 
+\vartheta \left[\begin{array}{cc} 0 \ 1 \\ 0 \ 1 \\
\end{array}\right](y,z)\ 
\vartheta \left[\begin{array}{cc} 1 \ 0 \\ 1 \ 0 \\
\end{array}\right](y,z)\ 
\vartheta \left[\begin{array}{cc} 0 \ 1 \\ 0 \ 1 \\
\end{array}\right](y',z')\ 
\vartheta \left[\begin{array}{cc} 1 \ 0 \\ 1 \ 0 \\
\end{array}\right](y',z')\ 
\nonumber\\
&&\hskip -15mm 
-\vartheta \left[\begin{array}{cc} 1 \ 1 \\ 0 \ 1 \\
\end{array}\right](y,z)\ 
\vartheta \left[\begin{array}{cc} 0 \ 0 \\ 1 \ 0 \\
\end{array}\right](y,z)\ 
\vartheta \left[\begin{array}{cc} 1 \ 1 \\ 0 \ 1 \\
\end{array}\right](y',z')\ 
\vartheta \left[\begin{array}{cc} 0 \ 0 \\ 1 \ 0 \\
\end{array}\right](y',z') . 
\label{e2-15}
\end{eqnarray} 
\noindent \underline{Addition Formula 9):}\\ 
\noindent Putting $\alpha=\tau_1/2+\tau_{12}/2+1/2$, 
$\beta=\tau_2/2+\tau_{12}/2+1/2$ in Eq.(\ref{e2-5}), 
we have
\begin{eqnarray}
&&\hskip -15mm  
\vartheta\left[\begin{array}{cc} 1 \ 1 \\ 1 \ 1 \\ 
\end{array}\right](0,0)\ 
\vartheta\left[\begin{array}{cc} 0 \ 0 \\ 1 \ 1 \\ 
\end{array}\right](0,0)\
\vartheta\left[\begin{array}{cc} 1 \ 1 \\ 1 \ 1 \\ 
\end{array}\right](y+y',z+z')\ 
\vartheta\left[\begin{array}{cc} 0 \ 0 \\ 1 \ 1 \\ 
\end{array}\right](y-y',z-z')
\nonumber\\
&&\hskip -15mm 
=\vartheta \left[\begin{array}{cc} 0 \ 0 \\ 1 \ 1 \\
\end{array}\right](y,z)\ 
\vartheta \left[\begin{array}{cc} 1 \ 1 \\ 1 \ 1 \\
\end{array}\right](y,z)\ 
\vartheta \left[\begin{array}{cc} 0 \ 0 \\ 1 \ 1 \\
\end{array}\right](y',z')\ 
\vartheta \left[\begin{array}{cc} 1 \ 1 \\ 1 \ 1 \\
\end{array}\right](y',z')\ 
\nonumber\\
&&\hskip -15mm 
-\vartheta \left[\begin{array}{cc} 1 \ 0 \\ 1 \ 1 \\
\end{array}\right](y,z)\ 
\vartheta \left[\begin{array}{cc} 0 \ 1 \\ 1 \ 1 \\
\end{array}\right](y,z)\ 
\vartheta \left[\begin{array}{cc} 1 \ 0 \\ 1 \ 1 \\
\end{array}\right](y',z')\ 
\vartheta \left[\begin{array}{cc} 0 \ 1 \\ 1 \ 1 \\
\end{array}\right](y',z')\ 
\nonumber\\
&&\hskip -15mm 
+\vartheta \left[\begin{array}{cc} 0 \ 1 \\ 0 \ 1 \\
\end{array}\right](y,z)\ 
\vartheta \left[\begin{array}{cc} 1 \ 0 \\ 0 \ 1 \\
\end{array}\right](y,z)\ 
\vartheta \left[\begin{array}{cc} 0 \ 1 \\ 0 \ 1 \\
\end{array}\right](y',z')\ 
\vartheta \left[\begin{array}{cc} 1 \ 0 \\ 0 \ 1 \\
\end{array}\right](y',z')\ 
\nonumber\\
&&\hskip -15mm 
-\vartheta \left[\begin{array}{cc} 1 \ 1 \\ 0 \ 1 \\
\end{array}\right](y,z)\ 
\vartheta \left[\begin{array}{cc} 0 \ 0 \\ 0 \ 1 \\
\end{array}\right](y,z)\ 
\vartheta \left[\begin{array}{cc} 1 \ 1 \\ 0 \ 1 \\
\end{array}\right](y',z')\ 
\vartheta \left[\begin{array}{cc} 0 \ 0 \\ 0 \ 1 \\
\end{array}\right](y',z')  . 
\label{e2-16}
\end{eqnarray} 
\noindent \underline{Addition Formula 10):}\\ 
\noindent Putting $\alpha=0$, $\beta=0$ in Eq.(\ref{e2-5}), 
we have
\begin{eqnarray}
&&\hskip -15mm  
\vartheta\left[\begin{array}{cc} 0 \ 0 \\ 0 \ 0 \\ 
\end{array}\right](0,0)\ 
\vartheta\left[\begin{array}{cc} 0 \ 0 \\ 1 \ 1 \\ 
\end{array}\right](0,0)\
\vartheta\left[\begin{array}{cc} 0 \ 0 \\ 0 \ 0 \\ 
\end{array}\right](y+y',z+z')\ 
\vartheta\left[\begin{array}{cc} 0 \ 0 \\ 1 \ 1 \\ 
\end{array}\right](y-y',z-z')
\nonumber\\
&&\hskip -15mm 
=\vartheta \left[\begin{array}{cc} 0 \ 0 \\ 1 \ 1 \\
\end{array}\right](y,z)\ 
\vartheta \left[\begin{array}{cc} 0 \ 0 \\ 0 \ 0 \\
\end{array}\right](y,z)\ 
\vartheta \left[\begin{array}{cc} 0 \ 0 \\ 1 \ 1 \\
\end{array}\right](y',z')\ 
\vartheta \left[\begin{array}{cc} 0 \ 0 \\ 0 \ 0 \\
\end{array}\right](y',z')\ 
\nonumber\\
&&\hskip -15mm 
-\vartheta \left[\begin{array}{cc} 1 \ 0 \\ 1 \ 1 \\
\end{array}\right](y,z)\ 
\vartheta \left[\begin{array}{cc} 1 \ 0 \\ 0 \ 0 \\
\end{array}\right](y,z)\ 
\vartheta \left[\begin{array}{cc} 1 \ 0 \\ 1 \ 1 \\
\end{array}\right](y',z')\ 
\vartheta \left[\begin{array}{cc} 1 \ 0 \\ 0 \ 0 \\
\end{array}\right](y',z')\ 
\nonumber\\
&&\hskip -15mm 
-\vartheta \left[\begin{array}{cc} 0 \ 1 \\ 0 \ 1 \\
\end{array}\right](y,z)\ 
\vartheta \left[\begin{array}{cc} 0 \ 1 \\ 1 \ 0 \\
\end{array}\right](y,z)\ 
\vartheta \left[\begin{array}{cc} 0 \ 1 \\ 0 \ 1 \\
\end{array}\right](y',z')\ 
\vartheta \left[\begin{array}{cc} 0 \ 1 \\ 1 \ 0 \\
\end{array}\right](y',z')\ 
\nonumber\\
&&\hskip -15mm 
+\vartheta \left[\begin{array}{cc} 1 \ 1 \\ 0 \ 1 \\
\end{array}\right](y,z)\ 
\vartheta \left[\begin{array}{cc} 1 \ 1 \\ 1 \ 0 \\
\end{array}\right](y,z)\ 
\vartheta \left[\begin{array}{cc} 1 \ 1 \\ 0 \ 1 \\
\end{array}\right](y',z')\ 
\vartheta \left[\begin{array}{cc} 1 \ 1 \\ 1 \ 0 \\
\end{array}\right](y',z')  . 
\label{e2-17}
\end{eqnarray} 
\noindent \underline{Addition Formula 11):}\\ 
\noindent Putting $\alpha=\tau_1/2+1/2$, $\beta=\tau_{12}/2+1/2$, which gives 
$\displaystyle{\vartheta\left[\begin{array}{cc} 0 \ 0 \\ 0 \ 0 \\ 
\end{array}\right](\alpha,\beta)=0}$, in Eq.(\ref{e2-6}), 
we have
\begin{eqnarray}
&&\hskip -15mm  
\vartheta\left[\begin{array}{cc} 1 \ 0 \\ 0 \ 1 \\ 
\end{array}\right](0,0)\ 
\vartheta\left[\begin{array}{cc} 0 \ 0 \\ 0 \ 1 \\ 
\end{array}\right](0,0)\
\vartheta\left[\begin{array}{cc} 1 \ 0 \\ 1 \ 1 \\ 
\end{array}\right](y+y',z+z')\ 
\vartheta\left[\begin{array}{cc} 0 \ 0 \\ 1 \ 1 \\ 
\end{array}\right](y-y',z-z')
\nonumber\\
&&\hskip -15mm 
=\vartheta \left[\begin{array}{cc} 0 \ 0 \\ 1 \ 1 \\
\end{array}\right](y,z)\ 
\vartheta \left[\begin{array}{cc} 1 \ 0 \\ 1 \ 1 \\
\end{array}\right](y,z)\ 
\vartheta \left[\begin{array}{cc} 0 \ 0 \\ 0 \ 1 \\
\end{array}\right](y',z')\ 
\vartheta \left[\begin{array}{cc} 1 \ 0 \\ 0 \ 1 \\
\end{array}\right](y',z')\ 
\nonumber\\
&&\hskip -15mm 
-\vartheta \left[\begin{array}{cc} 0 \ 1 \\ 1 \ 1 \\
\end{array}\right](y,z)\ 
\vartheta \left[\begin{array}{cc} 1 \ 1 \\ 1 \ 1 \\
\end{array}\right](y,z)\ 
\vartheta \left[\begin{array}{cc} 0 \ 1 \\ 0 \ 1 \\
\end{array}\right](y',z')\ 
\vartheta \left[\begin{array}{cc} 1 \ 1 \\ 0 \ 1 \\
\end{array}\right](y',z')\ 
\nonumber\\
&&\hskip -15mm 
+\vartheta \left[\begin{array}{cc} 0 \ 0 \\ 0 \ 1 \\
\end{array}\right](y,z)\ 
\vartheta \left[\begin{array}{cc} 1 \ 0 \\ 0 \ 1 \\
\end{array}\right](y,z)\ 
\vartheta \left[\begin{array}{cc} 0 \ 0 \\ 1 \ 1 \\
\end{array}\right](y',z')\ 
\vartheta \left[\begin{array}{cc} 1 \ 0 \\ 1 \ 1 \\
\end{array}\right](y',z')\ 
\nonumber\\
&&\hskip -15mm 
-\vartheta \left[\begin{array}{cc} 0 \ 1 \\ 0 \ 1 \\
\end{array}\right](y,z)\ 
\vartheta \left[\begin{array}{cc} 1 \ 1 \\ 0 \ 1 \\
\end{array}\right](y,z)\ 
\vartheta \left[\begin{array}{cc} 0 \ 1 \\ 1 \ 1 \\
\end{array}\right](y',z')\ 
\vartheta \left[\begin{array}{cc} 1 \ 1 \\ 1 \ 1 \\
\end{array}\right](y',z')  . 
\label{e2-18}
\end{eqnarray} 
\noindent \underline{Addition Formula 12):}\\ 
\noindent Putting $\alpha=\tau_1/2+1/2$, $\beta=\tau_{12}/2$, which gives 
$\displaystyle{\vartheta\left[\begin{array}{cc} 0 \ 0 \\ 0 \ 0 \\ 
\end{array}\right](\alpha,\beta)=0}$, in Eq.(\ref{e2-6}), 
we have
\begin{eqnarray}
&&\hskip -15mm  
\vartheta\left[\begin{array}{cc} 1 \ 0 \\ 0 \ 0 \\ 
\end{array}\right](0,0)\ 
\vartheta\left[\begin{array}{cc} 0 \ 0 \\ 0 \ 1 \\ 
\end{array}\right](0,0)\
\vartheta\left[\begin{array}{cc} 1 \ 0 \\ 1 \ 0 \\ 
\end{array}\right](y+y',z+z')\ 
\vartheta\left[\begin{array}{cc} 0 \ 0 \\ 1 \ 1 \\ 
\end{array}\right](y-y',z-z')
\nonumber\\
&&\hskip -15mm 
=\vartheta \left[\begin{array}{cc} 0 \ 0 \\ 1 \ 1 \\
\end{array}\right](y,z)\ 
\vartheta \left[\begin{array}{cc} 1 \ 0 \\ 1 \ 0 \\
\end{array}\right](y,z)\ 
\vartheta \left[\begin{array}{cc} 0 \ 0 \\ 0 \ 1 \\
\end{array}\right](y',z')\ 
\vartheta \left[\begin{array}{cc} 1 \ 0 \\ 0 \ 0 \\
\end{array}\right](y',z')\ 
\nonumber\\
&&\hskip -15mm 
-\vartheta \left[\begin{array}{cc} 0 \ 1 \\ 1 \ 1 \\
\end{array}\right](y,z)\ 
\vartheta \left[\begin{array}{cc} 1 \ 1 \\ 1 \ 0 \\
\end{array}\right](y,z)\ 
\vartheta \left[\begin{array}{cc} 0 \ 1 \\ 0 \ 1 \\
\end{array}\right](y',z')\ 
\vartheta \left[\begin{array}{cc} 1 \ 1 \\ 0 \ 0 \\
\end{array}\right](y',z')\ 
\nonumber\\
&&\hskip -15mm 
+\vartheta \left[\begin{array}{cc} 0 \ 0 \\ 0 \ 1 \\
\end{array}\right](y,z)\ 
\vartheta \left[\begin{array}{cc} 1 \ 0 \\ 0 \ 0 \\
\end{array}\right](y,z)\ 
\vartheta \left[\begin{array}{cc} 0 \ 0 \\ 1 \ 1 \\
\end{array}\right](y',z')\ 
\vartheta \left[\begin{array}{cc} 1 \ 0 \\ 1 \ 0 \\
\end{array}\right](y',z')\ 
\nonumber\\
&&\hskip -15mm 
-\vartheta \left[\begin{array}{cc} 0 \ 1 \\ 0 \ 1 \\
\end{array}\right](y,z)\ 
\vartheta \left[\begin{array}{cc} 1 \ 1 \\ 0 \ 0 \\
\end{array}\right](y,z)\ 
\vartheta \left[\begin{array}{cc} 0 \ 1 \\ 1 \ 1 \\
\end{array}\right](y',z')\ 
\vartheta \left[\begin{array}{cc} 1 \ 1 \\ 1 \ 0 \\
\end{array}\right](y',z') . 
\label{e2-19}
\end{eqnarray} 
\noindent \underline{Addition Formula 13):}\\ 
\noindent Putting $\alpha=\tau_1/2+\tau_{12}/2+1/2$, 
$\beta=\tau_2/2+\tau_{12}/2$, which gives 
$\displaystyle{\vartheta\left[\begin{array}{cc} 0 \ 0 \\ 0 \ 0 \\ 
\end{array}\right](\alpha,\beta)=0}$, in Eq.(\ref{e2-6}), 
we have
\begin{eqnarray}
&&\hskip -15mm  
\vartheta\left[\begin{array}{cc} 1 \ 1 \\ 0 \ 0 \\ 
\end{array}\right](0,0)\ 
\vartheta\left[\begin{array}{cc} 0 \ 0 \\ 0 \ 1 \\ 
\end{array}\right](0,0)\
\vartheta\left[\begin{array}{cc} 1 \ 1 \\ 1 \ 0 \\ 
\end{array}\right](y+y',z+z')\ 
\vartheta\left[\begin{array}{cc} 0 \ 0 \\ 1 \ 1 \\ 
\end{array}\right](y-y',z-z')
\nonumber\\
&&\hskip -15mm 
=\vartheta \left[\begin{array}{cc} 0 \ 0 \\ 1 \ 1 \\
\end{array}\right](y,z)\ 
\vartheta \left[\begin{array}{cc} 1 \ 1 \\ 1 \ 0 \\
\end{array}\right](y,z)\ 
\vartheta \left[\begin{array}{cc} 0 \ 0 \\ 0 \ 1 \\
\end{array}\right](y',z')\ 
\vartheta \left[\begin{array}{cc} 1 \ 1 \\ 0 \ 0 \\
\end{array}\right](y',z')\ 
\nonumber\\
&&\hskip -15mm 
-\vartheta \left[\begin{array}{cc} 0 \ 1 \\ 1 \ 1 \\
\end{array}\right](y,z)\ 
\vartheta \left[\begin{array}{cc} 1 \ 0 \\ 1 \ 0 \\
\end{array}\right](y,z)\ 
\vartheta \left[\begin{array}{cc} 0 \ 1 \\ 0 \ 1 \\
\end{array}\right](y',z')\ 
\vartheta \left[\begin{array}{cc} 1 \ 0 \\ 0 \ 0 \\
\end{array}\right](y',z')\ 
\nonumber\\
&&\hskip -15mm 
+\vartheta \left[\begin{array}{cc} 0 \ 0 \\ 0 \ 1 \\
\end{array}\right](y,z)\ 
\vartheta \left[\begin{array}{cc} 1 \ 1 \\ 0 \ 0 \\
\end{array}\right](y,z)\ 
\vartheta \left[\begin{array}{cc} 0 \ 0 \\ 1 \ 1 \\
\end{array}\right](y',z')\ 
\vartheta \left[\begin{array}{cc} 1 \ 1 \\ 1 \ 0 \\
\end{array}\right](y',z')\ 
\nonumber\\
&&\hskip -15mm 
-\vartheta \left[\begin{array}{cc} 0 \ 1 \\ 0 \ 1 \\
\end{array}\right](y,z)\ 
\vartheta \left[\begin{array}{cc} 1 \ 0 \\ 0 \ 0 \\
\end{array}\right](y,z)\ 
\vartheta \left[\begin{array}{cc} 0 \ 1 \\ 1 \ 1 \\
\end{array}\right](y',z')\ 
\vartheta \left[\begin{array}{cc} 1 \ 0 \\ 1 \ 0 \\
\end{array}\right](y',z')  .  
\label{e2-20}
\end{eqnarray} 
\noindent \underline{Addition Formula 14):}\\ 
\noindent Putting $\alpha=\tau_{12}/2+1/2$, 
$\beta=\tau_2/2+1/2$, which gives 
$\displaystyle{\vartheta\left[\begin{array}{cc} 0 \ 0 \\ 0 \ 0 \\ 
\end{array}\right](\alpha,\beta)=0}$, in Eq.(\ref{e2-7}), 
we have
\begin{eqnarray}
&&\hskip -15mm  
\vartheta\left[\begin{array}{cc} 0 \ 1 \\ 1 \ 0 \\ 
\end{array}\right](0,0)\ 
\vartheta\left[\begin{array}{cc} 0 \ 0 \\ 1 \ 0 \\ 
\end{array}\right](0,0)\
\vartheta\left[\begin{array}{cc} 0 \ 1 \\ 1 \ 1 \\ 
\end{array}\right](y+y',z+z')\ 
\vartheta\left[\begin{array}{cc} 0 \ 0 \\ 1 \ 1 \\ 
\end{array}\right](y-y',z-z')
\nonumber\\
&&\hskip -15mm 
=\vartheta \left[\begin{array}{cc} 0 \ 0 \\ 1 \ 1 \\
\end{array}\right](y,z)\ 
\vartheta \left[\begin{array}{cc} 0 \ 1 \\ 1 \ 1 \\
\end{array}\right](y,z)\ 
\vartheta \left[\begin{array}{cc} 0 \ 0 \\ 1 \ 0 \\
\end{array}\right](y',z')\ 
\vartheta \left[\begin{array}{cc} 0 \ 1 \\ 1 \ 0 \\
\end{array}\right](y',z')\ 
\nonumber\\
&&\hskip -15mm 
-\vartheta \left[\begin{array}{cc} 1 \ 0 \\ 1 \ 1 \\
\end{array}\right](y,z)\ 
\vartheta \left[\begin{array}{cc} 1 \ 1 \\ 1 \ 1 \\
\end{array}\right](y,z)\ 
\vartheta \left[\begin{array}{cc} 1 \ 0 \\ 1 \ 0 \\
\end{array}\right](y',z')\ 
\vartheta \left[\begin{array}{cc} 1 \ 1 \\ 1 \ 0 \\
\end{array}\right](y',z')\ 
\nonumber\\
&&\hskip -15mm 
+\vartheta \left[\begin{array}{cc} 0 \ 0 \\ 1 \ 0 \\
\end{array}\right](y,z)\ 
\vartheta \left[\begin{array}{cc} 0 \ 1 \\ 1 \ 0 \\
\end{array}\right](y,z)\ 
\vartheta \left[\begin{array}{cc} 0 \ 0 \\ 1 \ 1 \\
\end{array}\right](y',z')\ 
\vartheta \left[\begin{array}{cc} 0 \ 1 \\ 1 \ 1 \\
\end{array}\right](y',z')\ 
\nonumber\\
&&\hskip -15mm 
-\vartheta \left[\begin{array}{cc} 1 \ 0 \\ 1 \ 0 \\
\end{array}\right](y,z)\ 
\vartheta \left[\begin{array}{cc} 1 \ 1 \\ 1 \ 0 \\
\end{array}\right](y,z)\ 
\vartheta \left[\begin{array}{cc} 1 \ 0 \\ 1 \ 1 \\
\end{array}\right](y',z')\ 
\vartheta \left[\begin{array}{cc} 1 \ 1 \\ 1 \ 1 \\
\end{array}\right](y',z')  . 
\label{e2-21}
\end{eqnarray} 
\noindent \underline{Addition Formula 15):}\\ 
\noindent Putting $\alpha=\tau_{12}/2$, 
$\beta=\tau_2/2+1/2$, which gives 
$\displaystyle{\vartheta\left[\begin{array}{cc} 0 \ 0 \\ 0 \ 0 \\ 
\end{array}\right](\alpha,\beta)=0}$, in Eq.(\ref{e2-7}), 
we have
\begin{eqnarray}
&&\hskip -15mm  
\vartheta\left[\begin{array}{cc} 0 \ 1 \\ 0 \ 0 \\ 
\end{array}\right](0,0)\ 
\vartheta\left[\begin{array}{cc} 0 \ 0 \\ 1 \ 0 \\ 
\end{array}\right](0,0)\
\vartheta\left[\begin{array}{cc} 0 \ 1 \\ 0 \ 1 \\ 
\end{array}\right](y+y',z+z')\ 
\vartheta\left[\begin{array}{cc} 0 \ 0 \\ 1 \ 1 \\ 
\end{array}\right](y-y',z-z')
\nonumber\\
&&\hskip -15mm 
=\vartheta \left[\begin{array}{cc} 0 \ 0 \\ 1 \ 1 \\
\end{array}\right](y,z)\ 
\vartheta \left[\begin{array}{cc} 0 \ 1 \\ 0 \ 1 \\
\end{array}\right](y,z)\ 
\vartheta \left[\begin{array}{cc} 0 \ 0 \\ 1 \ 0 \\
\end{array}\right](y',z')\ 
\vartheta \left[\begin{array}{cc} 0 \ 1 \\ 0 \ 0 \\
\end{array}\right](y',z')\ 
\nonumber\\
&&\hskip -15mm 
-\vartheta \left[\begin{array}{cc} 1 \ 0 \\ 1 \ 1 \\
\end{array}\right](y,z)\ 
\vartheta \left[\begin{array}{cc} 1 \ 1 \\ 0 \ 1 \\
\end{array}\right](y,z)\ 
\vartheta \left[\begin{array}{cc} 1 \ 0 \\ 1 \ 0 \\
\end{array}\right](y',z')\ 
\vartheta \left[\begin{array}{cc} 1 \ 1 \\ 0 \ 0 \\
\end{array}\right](y',z')\ 
\nonumber\\
&&\hskip -15mm 
+\vartheta \left[\begin{array}{cc} 0 \ 0 \\ 1 \ 0 \\
\end{array}\right](y,z)\ 
\vartheta \left[\begin{array}{cc} 0 \ 1 \\ 0 \ 0 \\
\end{array}\right](y,z)\ 
\vartheta \left[\begin{array}{cc} 0 \ 0 \\ 1 \ 1 \\
\end{array}\right](y',z')\ 
\vartheta \left[\begin{array}{cc} 0 \ 1 \\ 0 \ 1 \\
\end{array}\right](y',z')\ 
\nonumber\\
&&\hskip -15mm 
-\vartheta \left[\begin{array}{cc} 1 \ 0 \\ 1 \ 0 \\
\end{array}\right](y,z)\ 
\vartheta \left[\begin{array}{cc} 1 \ 1 \\ 0 \ 0 \\
\end{array}\right](y,z)\ 
\vartheta \left[\begin{array}{cc} 1 \ 0 \\ 1 \ 1 \\
\end{array}\right](y',z')\ 
\vartheta \left[\begin{array}{cc} 1 \ 1 \\ 0 \ 1 \\
\end{array}\right](y',z') . 
\label{e2-22}
\end{eqnarray} 
\noindent \underline{Addition Formula 16):}\\ 
\noindent Putting $\alpha=\tau_1/2+\tau_{12}/2$, 
$\beta=\tau_2/2+\tau_{12}/2+1/2$, which gives 
$\displaystyle{\vartheta\left[\begin{array}{cc} 0 \ 0 \\ 0 \ 0 \\ 
\end{array}\right](\alpha,\beta)=0}$, in Eq.(\ref{e2-7}), 
we have
\begin{eqnarray}
&&\hskip -15mm  
\vartheta\left[\begin{array}{cc} 1 \ 1 \\ 0 \ 0 \\ 
\end{array}\right](0,0)\ 
\vartheta\left[\begin{array}{cc} 0 \ 0 \\ 1 \ 0 \\ 
\end{array}\right](0,0)\
\vartheta\left[\begin{array}{cc} 1 \ 1 \\ 0 \ 1 \\ 
\end{array}\right](y+y',z+z')\ 
\vartheta\left[\begin{array}{cc} 0 \ 0 \\ 1 \ 1 \\ 
\end{array}\right](y-y',z-z')
\nonumber\\
&&\hskip -15mm 
=\vartheta \left[\begin{array}{cc} 0 \ 0 \\ 1 \ 1 \\
\end{array}\right](y,z)\ 
\vartheta \left[\begin{array}{cc} 1 \ 1 \\ 0 \ 1 \\
\end{array}\right](y,z)\ 
\vartheta \left[\begin{array}{cc} 0 \ 0 \\ 1 \ 0 \\
\end{array}\right](y',z')\ 
\vartheta \left[\begin{array}{cc} 1 \ 1 \\ 0 \ 0 \\
\end{array}\right](y',z')\ 
\nonumber\\
&&\hskip -15mm 
-\vartheta \left[\begin{array}{cc} 1 \ 0 \\ 1 \ 1 \\
\end{array}\right](y,z)\ 
\vartheta \left[\begin{array}{cc} 0 \ 1 \\ 0 \ 1 \\
\end{array}\right](y,z)\ 
\vartheta \left[\begin{array}{cc} 1 \ 0 \\ 1 \ 0 \\
\end{array}\right](y',z')\ 
\vartheta \left[\begin{array}{cc} 0 \ 1 \\ 0 \ 0 \\
\end{array}\right](y',z')\ 
\nonumber\\
&&\hskip -15mm 
+\vartheta \left[\begin{array}{cc} 0 \ 0 \\ 1 \ 0 \\
\end{array}\right](y,z)\ 
\vartheta \left[\begin{array}{cc} 1 \ 1 \\ 0 \ 0 \\
\end{array}\right](y,z)\ 
\vartheta \left[\begin{array}{cc} 0 \ 0 \\ 1 \ 1 \\
\end{array}\right](y',z')\ 
\vartheta \left[\begin{array}{cc} 1 \ 1 \\ 0 \ 1 \\
\end{array}\right](y',z')\ 
\nonumber\\
&&\hskip -15mm 
-\vartheta \left[\begin{array}{cc} 1 \ 0 \\ 1 \ 0 \\
\end{array}\right](y,z)\ 
\vartheta \left[\begin{array}{cc} 0 \ 1 \\ 0 \ 0 \\
\end{array}\right](y,z)\ 
\vartheta \left[\begin{array}{cc} 1 \ 0 \\ 1 \ 1 \\
\end{array}\right](y',z')\ 
\vartheta \left[\begin{array}{cc} 0 \ 1 \\ 0 \ 1 \\
\end{array}\right](y',z')  . 
\label{e2-23}
\end{eqnarray} 
We have numerically checked these relations Eq.(\ref{e2-8}) $\sim$
Eq.(\ref{e2-23}) by REDUCE.

\vskip 10mm
\newpage
\setcounter{equation}{0}
\section{Addition Formula of $g=2$ Hyperelliptic Function}
We define the hyperelliptic function in the form
\begin{eqnarray}
&&F\left[\begin{array}{cc} a \ c \\ b \ d \\
\end{array}\right](y,z)
=\vartheta \left[\begin{array}{cc} a \ c \\ b \ d \\
\end{array}\right](y,z)/
\vartheta \left[\begin{array}{cc} 0 \ 0 \\ 1 \ 1 \\
\end{array}\right](y,z) . 
\label{e3-1}
\end{eqnarray}
\noindent \underline{Addition Formula 1):}\\
Taking the ratio of Eq.(\ref{e2-9})/Eq.(\ref{e2-8}), we have 
the addition formula
\begin{eqnarray}
&&F\left[\begin{array}{cc} 0 \ 0 \\ 0 \ 1 \\
\end{array}\right](y+y',z+z')=\frac{A_1}{B_0 B_1} , 
\label{e3-2}\\ 
\nonumber\\
&&{\rm where\ }
\nonumber\\
&&A_1=
F\left[\begin{array}{cc} 0 \ 0 \\ 0 \ 1 \\
\end{array}\right](y,z)
F\left[\begin{array}{cc} 0 \ 0 \\ 0 \ 1 \\
\end{array}\right](y',z')
\nonumber\\
&&-F\left[\begin{array}{cc} 1 \ 0 \\ 1 \ 1 \\
\end{array}\right](y,z)
F\left[\begin{array}{cc} 1 \ 0 \\ 0 \ 1 \\
\end{array}\right](y,z)
F\left[\begin{array}{cc} 1 \ 0 \\ 1 \ 1 \\
\end{array}\right](y',z')
F\left[\begin{array}{cc} 1 \ 0 \\ 0 \ 1 \\
\end{array}\right](y',z')
\nonumber\\
&&-F\left[\begin{array}{cc} 0 \ 1 \\ 0 \ 1 \\
\end{array}\right](y,z)
F\left[\begin{array}{cc} 0 \ 1 \\ 1 \ 1 \\
\end{array}\right](y,z)
F\left[\begin{array}{cc} 0 \ 1 \\ 0 \ 1 \\
\end{array}\right](y',z')
F\left[\begin{array}{cc} 0 \ 1 \\ 1 \ 1 \\
\end{array}\right](y',z')
\nonumber\\
&&+F\left[\begin{array}{cc} 1 \ 1 \\ 0 \ 1 \\
\end{array}\right](y,z)
F\left[\begin{array}{cc} 1 \ 1 \\ 1 \ 1 \\
\end{array}\right](y,z)
F\left[\begin{array}{cc} 1 \ 1 \\ 0 \ 1 \\
\end{array}\right](y',z')
F\left[\begin{array}{cc} 1 \ 1 \\ 1 \ 1 \\
\end{array}\right](y',z')   ,
\nonumber\\
&&B_0=1-F^2 \left[\begin{array}{cc} 1 \ 0 \\ 1 \ 1 \\
\end{array}\right](y,z)
F^2 \left[\begin{array}{cc} 1 \ 0 \\ 1 \ 1 \\
\end{array}\right](y',z')
-F^2 \left[\begin{array}{cc} 0 \ 1 \\ 0 \ 1 \\
\end{array}\right](y,z)
F^2 \left[\begin{array}{cc} 0 \ 1 \\ 0 \ 1 \\
\end{array}\right](y',z')
\nonumber\\
&&+F^2 \left[\begin{array}{cc} 1 \ 1 \\ 0 \ 1 \\
\end{array}\right](y,z)
F^2 \left[\begin{array}{cc} 1 \ 1 \\ 0 \ 1 \\
\end{array}\right](y',z')  ,
\nonumber\\
&&B_1=F\left[\begin{array}{cc} 0 \ 0 \\ 0 \ 1 \\
\end{array}\right](0,0)  . 
\nonumber
\end{eqnarray}
\noindent \underline{Addition Formula 2):}\\
Taking the ratio of Eq.(\ref{e2-10})/Eq.(\ref{e2-8}), we have 
the addition formula
\begin{eqnarray}
&&F\left[\begin{array}{cc} 0 \ 0 \\ 1 \ 0 \\
\end{array}\right](y+y',z+z')=\frac{A_2}{B_0 B_2}  , 
\label{e3-3}\\ 
&&{\rm where\ }\nonumber\\
&&A_2=
F\left[\begin{array}{cc} 0 \ 0 \\ 1 \ 0 \\
\end{array}\right](y,z)
F\left[\begin{array}{cc} 0 \ 0 \\ 1 \ 0 \\
\end{array}\right](y',z')
\nonumber\\
&&-F\left[\begin{array}{cc} 1 \ 0 \\ 1 \ 1 \\
\end{array}\right](y,z)
F\left[\begin{array}{cc} 1 \ 0 \\ 1 \ 0 \\
\end{array}\right](y,z)
F\left[\begin{array}{cc} 1 \ 0 \\ 1 \ 1 \\
\end{array}\right](y',z')
F\left[\begin{array}{cc} 1 \ 0 \\ 1 \ 0 \\
\end{array}\right](y',z')
\nonumber\\
&&-F\left[\begin{array}{cc} 0 \ 1 \\ 0 \ 1 \\
\end{array}\right](y,z)
F\left[\begin{array}{cc} 0 \ 1 \\ 0 \ 0 \\
\end{array}\right](y,z)
F\left[\begin{array}{cc} 0 \ 1 \\ 0 \ 1 \\
\end{array}\right](y',z')
F\left[\begin{array}{cc} 0 \ 1 \\ 0 \ 0 \\
\end{array}\right](y',z')
\nonumber\\
&&+F\left[\begin{array}{cc} 1 \ 1 \\ 0 \ 1 \\
\end{array}\right](y,z)
F\left[\begin{array}{cc} 1 \ 1 \\ 0 \ 0 \\
\end{array}\right](y,z)
F\left[\begin{array}{cc} 1 \ 1 \\ 0 \ 1 \\
\end{array}\right](y',z')
F\left[\begin{array}{cc} 1 \ 1 \\ 0 \ 0 \\
\end{array}\right](y',z')   , 
\nonumber\\
&&B_2=F\left[\begin{array}{cc} 0 \ 0 \\ 1 \ 0 \\
\end{array}\right](0,0) . 
\nonumber
\end{eqnarray}
\noindent \underline{Addition Formula 3):}\\
Taking the ratio of Eq.(\ref{e2-11})/Eq.(\ref{e2-8}), we have 
the addition formula
\begin{eqnarray}
&&F\left[\begin{array}{cc} 1 \ 0 \\ 0 \ 1 \\
\end{array}\right](y+y',z+z')=\frac{A_3}{B_0 B_3}  , 
\label{e3-4}\\ 
&&{\rm where\ }\nonumber\\
&&A_3=
F\left[\begin{array}{cc} 1 \ 0 \\ 0 \ 1 \\
\end{array}\right](y,z)
F\left[\begin{array}{cc} 1 \ 0 \\ 0 \ 1 \\
\end{array}\right](y',z')
\nonumber\\
&&-F\left[\begin{array}{cc} 1 \ 0 \\ 1 \ 1 \\
\end{array}\right](y,z)
F\left[\begin{array}{cc} 0 \ 0 \\ 0 \ 1 \\
\end{array}\right](y,z)
F\left[\begin{array}{cc} 1 \ 0 \\ 1 \ 1 \\
\end{array}\right](y',z')
F\left[\begin{array}{cc} 0 \ 0 \\ 0 \ 1 \\
\end{array}\right](y',z')
\nonumber\\
&&+F\left[\begin{array}{cc} 0 \ 1 \\ 0 \ 1 \\
\end{array}\right](y,z)
F\left[\begin{array}{cc} 1 \ 1 \\ 1 \ 1 \\
\end{array}\right](y,z)
F\left[\begin{array}{cc} 0 \ 1 \\ 0 \ 1 \\
\end{array}\right](y',z')
F\left[\begin{array}{cc} 1 \ 1 \\ 1 \ 1 \\
\end{array}\right](y',z')
\nonumber\\
&&-F\left[\begin{array}{cc} 1 \ 1 \\ 0 \ 1 \\
\end{array}\right](y,z)
F\left[\begin{array}{cc} 0 \ 1 \\ 1 \ 1 \\
\end{array}\right](y,z)
F\left[\begin{array}{cc} 1 \ 1 \\ 0 \ 1 \\
\end{array}\right](y',z')
F\left[\begin{array}{cc} 0 \ 1 \\ 1 \ 1 \\
\end{array}\right](y',z')   , 
\nonumber\\
&&B_3=F\left[\begin{array}{cc} 1 \ 0 \\ 0 \ 1 \\
\end{array}\right](0,0) . 
\nonumber
\end{eqnarray}
\noindent \underline{Addition Formula 4):}\\
Taking the ratio of Eq.(\ref{e2-12})/Eq.(\ref{e2-8}), we have 
the addition formula
\begin{eqnarray}
&&F\left[\begin{array}{cc} 0 \ 1 \\ 1 \ 0 \\
\end{array}\right](y+y',z+z')=\frac{A_4}{B_0 B_4}  , 
\label{e3-5}\\ 
&&{\rm where\ }\nonumber\\
&&A_4=
F\left[\begin{array}{cc} 0 \ 1 \\ 1 \ 0 \\
\end{array}\right](y,z)
F\left[\begin{array}{cc} 0 \ 1 \\ 1 \ 0 \\
\end{array}\right](y',z')
\nonumber\\
&&-F\left[\begin{array}{cc} 1 \ 0 \\ 1 \ 1 \\
\end{array}\right](y,z)
F\left[\begin{array}{cc} 1 \ 1 \\ 1 \ 0 \\
\end{array}\right](y,z)
F\left[\begin{array}{cc} 1 \ 0 \\ 1 \ 1 \\
\end{array}\right](y',z')
F\left[\begin{array}{cc} 1 \ 1 \\ 1 \ 0 \\
\end{array}\right](y',z')
\nonumber\\
&&-F\left[\begin{array}{cc} 0 \ 1 \\ 0 \ 1 \\
\end{array}\right](y,z)
F\left[\begin{array}{cc} 0 \ 0 \\ 0 \ 0 \\
\end{array}\right](y,z)
F\left[\begin{array}{cc} 0 \ 1 \\ 0 \ 1 \\
\end{array}\right](y',z')
F\left[\begin{array}{cc} 0 \ 0 \\ 0 \ 0 \\
\end{array}\right](y',z')
\nonumber\\
&&+F\left[\begin{array}{cc} 1 \ 1 \\ 0 \ 1 \\
\end{array}\right](y,z)
F\left[\begin{array}{cc} 1 \ 0 \\ 0 \ 0 \\
\end{array}\right](y,z)
F\left[\begin{array}{cc} 1 \ 1 \\ 0 \ 1 \\
\end{array}\right](y',z')
F\left[\begin{array}{cc} 1 \ 0 \\ 0 \ 0 \\
\end{array}\right](y',z')   , 
\nonumber\\
&&B_4=F\left[\begin{array}{cc} 0 \ 1 \\ 1 \ 0 \\
\end{array}\right](0,0) . 
\nonumber
\end{eqnarray}
\noindent \underline{Addition Formula 5):}\\
Taking the ratio of Eq.(\ref{e2-13})/Eq.(\ref{e2-8}), we have 
the addition formula
\begin{eqnarray}
&&F\left[\begin{array}{cc} 0 \ 1 \\ 0 \ 0 \\
\end{array}\right](y+y',z+z')=\frac{A_5}{B_0 B_5}  , 
\label{e3-6}\\ 
&&{\rm where\ }\nonumber\\
&&A_5=
F\left[\begin{array}{cc} 0 \ 1 \\ 0 \ 0 \\
\end{array}\right](y,z)
F\left[\begin{array}{cc} 0 \ 1 \\ 0 \ 0 \\
\end{array}\right](y',z')
\nonumber\\
&&-F\left[\begin{array}{cc} 1 \ 0 \\ 1 \ 1 \\
\end{array}\right](y,z)
F\left[\begin{array}{cc} 1 \ 1 \\ 0 \ 0 \\
\end{array}\right](y,z)
F\left[\begin{array}{cc} 1 \ 0 \\ 1 \ 1 \\
\end{array}\right](y',z')
F\left[\begin{array}{cc} 1 \ 1 \\ 0 \ 0 \\
\end{array}\right](y',z')
\nonumber\\
&&-F\left[\begin{array}{cc} 0 \ 1 \\ 0 \ 1 \\
\end{array}\right](y,z)
F\left[\begin{array}{cc} 0 \ 0 \\ 1 \ 0 \\
\end{array}\right](y,z)
F\left[\begin{array}{cc} 0 \ 1 \\ 0 \ 1 \\
\end{array}\right](y',z')
F\left[\begin{array}{cc} 0 \ 0 \\ 1 \ 0 \\
\end{array}\right](y',z')
\nonumber\\
&&+F\left[\begin{array}{cc} 1 \ 1 \\ 0 \ 1 \\
\end{array}\right](y,z)
F\left[\begin{array}{cc} 1 \ 0 \\ 1 \ 0 \\
\end{array}\right](y,z)
F\left[\begin{array}{cc} 1 \ 1 \\ 0 \ 1 \\
\end{array}\right](y',z')
F\left[\begin{array}{cc} 1 \ 0 \\ 1 \ 0 \\
\end{array}\right](y',z')   , 
\nonumber\\
&&B_5=F\left[\begin{array}{cc} 0 \ 1 \\ 0 \ 0 \\
\end{array}\right](0,0) . 
\nonumber
\end{eqnarray}
\noindent \underline{Addition Formula 6):}\\
Taking the ratio of Eq.(\ref{e2-14})/Eq.(\ref{e2-8}), we have 
the addition formula
\begin{eqnarray}
&&F\left[\begin{array}{cc} 1 \ 0 \\ 0 \ 0 \\
\end{array}\right](y+y',z+z')=\frac{A_6}{B_0 B_6}  , 
\label{e3-7}\\ 
&&{\rm where\ }\nonumber\\
&&A_6=
F\left[\begin{array}{cc} 1 \ 0 \\ 0 \ 0 \\
\end{array}\right](y,z)
F\left[\begin{array}{cc} 1 \ 0 \\ 0 \ 0 \\
\end{array}\right](y',z')
\nonumber\\
&&-F\left[\begin{array}{cc} 1 \ 0 \\ 1 \ 1 \\
\end{array}\right](y,z)
F\left[\begin{array}{cc} 0 \ 0 \\ 0 \ 0 \\
\end{array}\right](y,z)
F\left[\begin{array}{cc} 1 \ 0 \\ 1 \ 1 \\
\end{array}\right](y',z')
F\left[\begin{array}{cc} 0 \ 0 \\ 0 \ 0 \\
\end{array}\right](y',z')
\nonumber\\
&&+F\left[\begin{array}{cc} 0 \ 1 \\ 0 \ 1 \\
\end{array}\right](y,z)
F\left[\begin{array}{cc} 1 \ 1 \\ 1 \ 0 \\
\end{array}\right](y,z)
F\left[\begin{array}{cc} 0 \ 1 \\ 0 \ 1 \\
\end{array}\right](y',z')
F\left[\begin{array}{cc} 1 \ 1 \\ 1 \ 0 \\
\end{array}\right](y',z')
\nonumber\\
&&-F\left[\begin{array}{cc} 1 \ 1 \\ 0 \ 1 \\
\end{array}\right](y,z)
F\left[\begin{array}{cc} 0 \ 1 \\ 1 \ 0 \\
\end{array}\right](y,z)
F\left[\begin{array}{cc} 1 \ 1 \\ 0 \ 1 \\
\end{array}\right](y',z')
F\left[\begin{array}{cc} 0 \ 1 \\ 1 \ 0 \\
\end{array}\right](y',z')   ,
\nonumber\\
&&B_6=F\left[\begin{array}{cc} 1 \ 0 \\ 0 \ 0 \\
\end{array}\right](0,0) . 
\nonumber
\end{eqnarray}
\noindent \underline{Addition Formula 7):}\\
Taking the ratio of Eq.(\ref{e2-15})/Eq.(\ref{e2-8}), we have 
the addition formula
\begin{eqnarray}
&&F\left[\begin{array}{cc} 1 \ 1 \\ 0 \ 0 \\
\end{array}\right](y+y',z+z')=\frac{A_7}{B_0 B_7}  , 
\label{e3-8}\\ 
&&{\rm where\ }\nonumber\\
&&A_7=
F\left[\begin{array}{cc} 1 \ 1 \\ 0 \ 0 \\
\end{array}\right](y,z)
F\left[\begin{array}{cc} 1 \ 1 \\ 0 \ 0 \\
\end{array}\right](y',z')
\nonumber\\
&&-F\left[\begin{array}{cc} 1 \ 0 \\ 1 \ 1 \\
\end{array}\right](y,z)
F\left[\begin{array}{cc} 0 \ 1 \\ 0 \ 0 \\
\end{array}\right](y,z)
F\left[\begin{array}{cc} 1 \ 0 \\ 1 \ 1 \\
\end{array}\right](y',z')
F\left[\begin{array}{cc} 0 \ 1 \\ 0 \ 0 \\
\end{array}%
\right](0,0) . 
\nonumber
\end{eqnarray}
We have numerically cheched these addition formula
Eq.(\ref{e3-2}) $\sim$ Eq.(\ref{e3-16}) by REDUCE.

\vskip 10mm
\newpage
\setcounter{equation}{0}
\section{Summary and Discussion}
The integrability condition of the Ising model is understood 
as the $SU(2)$ integrability condition and also as the model
parameterized by the elliptic function, where the integrability
condition is understood as the addition theorem of the elliptic
function.
The generalization of this integrability condition is 
to find the solution of the higher rank Lie group integrability
condition and also to find the model parameterized by higher genus
hyperelliptic function.
For the preparation of this purpose, we give the explicit formula
of the addition formula for $g=2$ hyperelliptic function.\\
The trivial case is the $SO(4)\cong [SU(2)\otimes SU(2)]/Z_2$ integrable
condition in the form
\begin{eqnarray}
  && U(x_1,y_1) V(x_1+x_2,y_1+y_2) U(x_2,y_2)
    =V(x_2,y_2)U(x_1+x_2,y_1+y_2) V(x_1,y_1) , 
\label{e4-1}\\
&&{\rm where}
\nonumber\\
&&U(x,y)=\tilde{U}(x,k_1)\otimes \tilde{U}(y,k_2),\quad 
V(x,y)=\tilde{V}(x,k_1)\otimes \tilde{V}(y,k_2) ,  
\label{e4-2}\\
&&{\rm and}
\nonumber\\
&&\tilde{U}(x,k_1)=\exp\{i{\rm am}(x,k_1) J_z\},\quad 
\tilde{U}(y,k_2)=\exp\{i{\rm am}(y,k_2) J_z\} ,
\nonumber\\
&&\tilde{V}(x,k_1)=\exp\{i{\rm am}(k_1 x,1/k_1) J_x\},\quad 
\tilde{V}(y,k_2)=\exp\{i{\rm am}(k_2 y,1/k_2) J_x\} . 
\nonumber 
\end{eqnarray}
The algebraic function with non-trivial two argument is 
$g=2$ hyperelliptic function instead of the direct product 
of the elliptic function. And we expect that the Lie group integrability 
condition will be written as the form of the $SO(5)\cong Sp(4,{\bf R})/Z_2$ 
Lie group integrable
condition.\\
In our case, two dimensional torus becomes the Jacobian varieties because
if we denote the period matrix as $\Omega$, we have
\begin{eqnarray}
&& ^t\Omega J \Omega=\left(\begin{array}{cc} 0 \ 0 \\ 0 \ 0\\
\end{array}\right)=0\ ,\quad 
\sqrt{-1}\ ^t\Omega J \bar{\Omega}=2 \left(\begin{array}{cc} 
{\rm Im}\tau_1 \ {\rm Im}\tau_{12} \\ {\rm Im}\tau_{12} \ {\rm Im}\tau_2\
\end{array}\right)\ >0 ,  
\label{e4-3}\\
&&{\rm where}
\nonumber\\
&& \Omega=\left(\begin{array}{cccc} 1 \ 0 \ \tau_1 \ \tau_{12} \\
0 \ 1 \ \tau_{12} \ \tau_2 \\
\end{array}\right)\ ,\quad 
J=\left(\begin{array}{cccc}  0 \ \ 0 \ \ 1 \ \ 0 \ \ \\  0 \ \ 0 \ \ 0 \ \ 1 \ \ \\
{-1} \ 0 \ \ 0 \ \ 0 \ \ \\  0 \  {-1} \ \ 0 \ \ 0 \ \ \\
\end{array}\right) , 
\nonumber
\end{eqnarray}
because we assume ${\rm Im}\tau_1>0$, ${\rm Im}\tau_2>0$, ${\rm Im}\tau_{12}>0$. \\
We conjecture that the addition formula for genus $g$ hyperelliptic function 
will be written as the integrable condition of the $Sp(2g,{\bf R})$ Lie group.
The reason of this conjecture is that we expect that the Lie group structure of the 
hyperelliptic function and the discrete $Sp(2g,{\bf Z})$ structure
of the moduli\cite{Siegel} will be mutually connected.Then the Jacobian 
varieties will have the $Sp(2g,{\bf R})$ Lie group structure through the addition 
formula of the theta function.\\
The more general conjecture is that the Abelian varieties will have the
general Lie group structure through the addition theorem of the theta function.\\
As the K3 surface can be patameterized by the $g=2$ hyperelliptic 
theta function\cite{Kumar}, some special algebraic varieties may have the 
Lie group structure.

\vskip 10mm
\newpage
\newpage

\newpage

\appendix
\setcounter{equation}{0}
\section{\large \bf Proof of the Fundamental Addition Relation of $g=2$ Theta Function}

\noindent \underline{Kossak's Fundamental Relation 1):}\\
If we take $\alpha'=1/2$, $\beta'=1/2$ in Eq.(\ref{e2-4}), which gives
$\vartheta\left[
\right](y',z')  . 
\nonumber\\
&& \label{A-1}
\end{eqnarray}
Then we replace $z \rightarrow z+1/2$, $z' \rightarrow z'+1/2$, and we have
\begin{eqnarray}
&&\hskip -20mm  \vartheta\left[\begin{array}{cc} 0 \ 0 \\ 0 \ 0 \\ 
\end{array}%
\right](y',z') .
\nonumber\\
&& \label{A-3}
\end{eqnarray}
In order to have another relation, we take another choice of 
$\alpha'=0$, $\beta'=1/2$ in Eq.(\ref{e2-4}),
which gives 
$\vartheta\left[%
\right](\alpha',\beta')=0$ , and we have 
\begin{eqnarray}
&&\hskip -20mm  \vartheta\left[%
\right](y',z')  . 
\nonumber\\
&& \label{A-4}
\end{eqnarray}
Then we replace $z\rightarrow z+1/2$, $z' \rightarrow z' +1/2$ in
Eq.(\ref{A-4}), we have
\begin{eqnarray}
&&\hskip -20mm  \vartheta\left[%
\right](y',z')  . 
\nonumber\\
&& \label{A-5}
\end{eqnarray}
The left-hand side of Eq.(\ref{A-4}) and Eq.(\ref{A-5}) is equal, 
so that we have the relation that the right-hand side of 
Eq.(\ref{A-4}) and Eq.(\ref{A-5}) is equal. Further, we replace
$\alpha \rightarrow \alpha+1/2$.
Then we have the relation
\begin{eqnarray}
&&\hskip -20mm  0=
\vartheta\left[%
\right](y',z')  . 
\label{A-7}
\end{eqnarray}
We further replace $y \rightarrow y+1/2$, $z \rightarrow z+1/2$, 
$y' \rightarrow y'+1/2$, $z' \rightarrow z'+1/2$, and we have
\begin{eqnarray}
&&\hskip -20mm  \vartheta\left[%
\right](y',z')  .  
\label{A-8} 
\end{eqnarray}
In the above, we use the following relation. First we notice 
$\vartheta\left[%
\right](v,u;\tau_2,\tau_1,\tau_{12})$, then the left-hand
side of Eq.(\ref{A-7}) does not change under the rename of 
$\alpha \leftrightarrow \beta$, $y \leftrightarrow z$, $y' \leftrightarrow z'$, 
$\tau_1 \leftrightarrow \tau_2$, then the first column and the second column
exchanged expression of the theta function in the right-hand side of Eq.(\ref{A-7})
is equal to the original expression.
This Eq.(\ref{A-8}) is the Kossak's Fundamental relation 1) of Eq.(\ref{e2-5}).\\
\noindent \underline{Kossak's Fundamental Relation 2):}\\
We choose $\alpha$, $\beta$\ (6 cases) in such a way as 
$\vartheta\left[\begin{array}{cc} 0 \ 0 \\ 0 \ 0 \\
\end{array}\right](\alpha,\beta)=0$.
Then Eq.(\ref{A-1}) and Eq.(\ref{A-2}) is given by
\begin{eqnarray}
&&\hskip -20mm \Big(\vartheta\left[\begin{array}{cc} 1 \ 1 \\ 0 \ 0 \\
\end{array}\right](\alpha,\beta)\ 
\vartheta\left[\begin{array}{cc} 1 \ 1 \\ 0 \ 0 \\
\end{array}\right](\frac{1}{2},\frac{1}{2})\ 
\vartheta\left[\begin{array}{cc} 1 \ 1 \\ 0 \ 0 \\
\end{array}\right](y+y'+\alpha,z+z'+\beta)\ 
\vartheta\left[\begin{array}{cc} 1 \ 1 \\ 0 \ 0 \\
\end{array}\right](y-y'+\frac{1}{2},z-z'+\frac{1}{2})\Big)
\nonumber\\
&&\hskip -20mm =\Big(\vartheta\left[\begin{array}{cc} 0 \ 0 \\ 0 \ 0 \\ 
\end{array}\right](y+\alpha+\frac{1}{2},z+\beta+\frac{1}{2})\ 
\vartheta\left[\begin{array}{cc} 0 \ 0 \\ 0 \ 0 \\ 
\end{array}\right](y,z)\ 
\vartheta\left[\begin{array}{cc} 0 \ 0 \\ 0 \ 0 \\ 
\end{array}\right](y'+\alpha+\frac{1}{2},z'+\beta+\frac{1}{2})\ 
\vartheta\left[\begin{array}{cc} 0 \ 0 \\ 0 \ 0 \\ 
\end{array}\right](y',z')\Big)
\nonumber\\
&&\hskip -20mm -\Big(\vartheta\left[\begin{array}{cc} 0 \ 1 \\ 0 \ 0 \\
\end{array}\right](y+\alpha+\frac{1}{2},z+\beta+\frac{1}{2})\ 
\vartheta\left[\begin{array}{cc} 0 \ 1 \\ 0 \ 0 \\
\end{array}\right](y,z)\  
\vartheta\left[\begin{array}{cc} 0 \ 1 \\ 0 \ 0 \\
\end{array}\right](y'+\alpha+\frac{1}{2},z'+\beta+\frac{1}{2})\  
\vartheta\left[\begin{array}{cc} 0 \ 1 \\ 0 \ 0 \\
\end{array}\right](y',z')\Big)
\nonumber\\
&&\hskip -20mm -\Big(\vartheta\left[\begin{array}{cc} 1 \ 0 \\ 0 \ 0 \\
\end{array}\right](y+\alpha+\frac{1}{2},z+\beta+\frac{1}{2})\  
\vartheta\left[\begin{array}{cc} 1 \ 0 \\ 0 \ 0 \\
\end{array}\right](y,z)\  
\vartheta\left[\begin{array}{cc} 1 \ 0 \\ 0 \ 0 \\
\end{array}\right](y'+\alpha+\frac{1}{2},z'+\beta+\frac{1}{2})\  
\vartheta\left[\begin{array}{cc} 1 \ 0 \\ 0 \ 0 \\
\end{array}\right](y',z')\Big)
\nonumber\\
&&\hskip -20mm +\Big(\vartheta\left[\begin{array}{cc} 1 \ 1 \\ 0 \ 0 \\
\end{array}\right](y+\alpha+\frac{1}{2},z+\beta+\frac{1}{2})\ 
\vartheta\left[\begin{array}{cc} 1 \ 1 \\ 0 \ 0 \\
\end{array}\right](y,z)\ 
\vartheta\left[\begin{array}{cc} 1 \ 1 \\ 0 \ 0 \\
\end{array}\right](y'+\alpha+\frac{1}{2},z'+\beta+\frac{1}{2})\ 
\vartheta\left[\begin{array}{cc} 1 \ 1 \\ 0 \ 0 \\
\end{array}\right](y',z')\Big)
\nonumber\\
&&\hskip -20mm =-\Big(\vartheta\left[\begin{array}{cc} 0 \ 0 \\ 0 \ 1 \\ 
\end{array}\right](y+\alpha+\frac{1}{2},z+\beta+\frac{1}{2})\ 
\vartheta\left[\begin{array}{cc} 0 \ 0 \\ 0 \ 1 \\ 
\end{array}\right](y,z)\ 
\vartheta\left[\begin{array}{cc} 0 \ 0 \\ 0 \ 1 \\ 
\end{array}\right](y'+\alpha+\frac{1}{2},z'+\beta+\frac{1}{2})\ 
\vartheta\left[\begin{array}{cc} 0 \ 0 \\ 0 \ 1 \\ 
\end{array}\right](y',z')\Big)
\nonumber\\
&&\hskip -20mm +\Big(\vartheta\left[\begin{array}{cc} 0 \ 1 \\ 0 \ 1 \\
\end{array}\right](y+\alpha+\frac{1}{2},z+\beta+\frac{1}{2})\ 
\vartheta\left[\begin{array}{cc} 0 \ 1 \\ 0 \ 1 \\
\end{array}\right](y,z)\  
\vartheta\left[\begin{array}{cc} 0 \ 1 \\ 0 \ 1 \\
\end{array}\right](y'+\alpha+\frac{1}{2},z'+\beta+\frac{1}{2})\  
\vartheta\left[\begin{array}{cc} 0 \ 1 \\ 0 \ 1 \\
\end{array}\right](y',z')\Big)
\nonumber\\
&&\hskip -20mm +\Big(\vartheta\left[\begin{array}{cc} 1 \ 0 \\ 0 \ 1 \\
\end{array}\right](y+\alpha+\frac{1}{2},z+\beta+\frac{1}{2})\  
\vartheta\left[\begin{array}{cc} 1 \ 0 \\ 0 \ 1 \\
\end{array}\right](y,z)\  
\vartheta\left[\begin{array}{cc} 1 \ 0 \\ 0 \ 1 \\
\end{array}\right](y'+\alpha+\frac{1}{2},z'+\beta+\frac{1}{2})\  
\vartheta\left[\begin{array}{cc} 1 \ 0 \\ 0 \ 1 \\
\end{array}\right](y',z')\Big)
\nonumber\\
&&\hskip -20mm -\Big(\vartheta\left[\begin{array}{cc} 1 \ 1 \\ 0 \ 1 \\
\end{array}\right](y+\alpha+\frac{1}{2},z+\beta+\frac{1}{2})\ 
\vartheta\left[\begin{array}{cc} 1 \ 1 \\ 0 \ 1 \\
\end{array}\right](y,z)\ 
\vartheta\left[\begin{array}{cc} 1 \ 1 \\ 0 \ 1 \\
\end{array}\right](y'+\alpha+\frac{1}{2},z'+\beta+\frac{1}{2})\ 
\vartheta\left[\begin{array}{cc} 1 \ 1 \\ 0 \ 1 \\
\end{array}\right](y',z')\Big)  . 
\nonumber\\
&& \label{A-9}
\end{eqnarray}
As the second term and the third term is equal, we have
\begin{eqnarray}
&&\hskip -20mm 0=\Big(\vartheta\left[\begin{array}{cc} 0 \ 0 \\ 0 \ 0 \\ 
\end{array}\right](y+\alpha+\frac{1}{2},z+\beta+\frac{1}{2})\ 
\vartheta\left[\begin{array}{cc} 0 \ 0 \\ 0 \ 0 \\ 
\end{array}\right](y,z)\ 
\vartheta\left[\begin{array}{cc} 0 \ 0 \\ 0 \ 0 \\ 
\end{array}\right](y'+\alpha+\frac{1}{2},z'+\beta+\frac{1}{2})\ 
\vartheta\left[\begin{array}{cc} 0 \ 0 \\ 0 \ 0 \\ 
\end{array}\right](y',z')\Big)
\nonumber\\
&&\hskip -20mm -\Big(\vartheta\left[\begin{array}{cc} 0 \ 1 \\ 0 \ 0 \\
\end{array}\right](y+\alpha+\frac{1}{2},z+\beta+\frac{1}{2})\ 
\vartheta\left[\begin{array}{cc} 0 \ 1 \\ 0 \ 0 \\
\end{array}\right](y,z)\  
\vartheta\left[\begin{array}{cc} 0 \ 1 \\ 0 \ 0 \\
\end{array}\right](y'+\alpha+\frac{1}{2},z'+\beta+\frac{1}{2})\  
\vartheta\left[\begin{array}{cc} 0 \ 1 \\ 0 \ 0 \\
\end{array}\right](y',z')\Big)
\nonumber\\
&&\hskip -20mm -\Big(\vartheta\left[\begin{array}{cc} 1 \ 0 \\ 0 \ 0 \\
\end{array}\right](y+\alpha+\frac{1}{2},z+\beta+\frac{1}{2})\  
\vartheta\left[\begin{array}{cc} 1 \ 0 \\ 0 \ 0 \\
\end{array}\right](y,z)\  
\vartheta\left[\begin{array}{cc} 1 \ 0 \\ 0 \ 0 \\
\end{array}\right](y'+\alpha+\frac{1}{2},z'+\beta+\frac{1}{2})\  
\vartheta\left[\begin{array}{cc} 1 \ 0 \\ 0 \ 0 \\
\end{array}\right](y',z')\Big)
\nonumber\\
&&\hskip -20mm +\Big(\vartheta\left[\begin{array}{cc} 1 \ 1 \\ 0 \ 0 \\
\end{array}\right](y+\alpha+\frac{1}{2},z+\beta+\frac{1}{2})\ 
\vartheta\left[\begin{array}{cc} 1 \ 1 \\ 0 \ 0 \\
\end{array}\right](y,z)\ 
\vartheta\left[\begin{array}{cc} 1 \ 1 \\ 0 \ 0 \\
\end{array}\right](y'+\alpha+\frac{1}{2},z'+\beta+\frac{1}{2})\ 
\vartheta\left[\begin{array}{cc} 1 \ 1 \\ 0 \ 0 \\
\end{array}\right](y',z')\Big)
\nonumber\\
&&\hskip -20mm +\Big(\vartheta\left[\begin{array}{cc} 0 \ 0 \\ 1 \ 0 \\ 
\end{array}\right](y+\alpha+\frac{1}{2},z+\beta+\frac{1}{2})\ 
\vartheta\left[\begin{array}{cc} 0 \ 0 \\ 1 \ 0  \\ 
\end{array}\right](y,z)\ 
\vartheta\left[\begin{array}{cc} 0 \ 0 \\ 1 \ 0  \\ 
\end{array}\right](y'+\alpha+\frac{1}{2},z'+\beta+\frac{1}{2})\ 
\vartheta\left[\begin{array}{cc} 0 \ 0 \\ 1 \ 0  \\ 
\end{array}\right](y',z')\Big)
\nonumber\\
&&\hskip -20mm -\Big(\vartheta\left[\begin{array}{cc} 1 \ 0 \\ 1 \ 0 \\
\end{array}\right](y+\alpha+\frac{1}{2},z+\beta+\frac{1}{2})\ 
\vartheta\left[\begin{array}{cc} 1 \ 0 \\ 1 \ 0 \\
\end{array}\right](y,z)\  
\vartheta\left[\begin{array}{cc} 1 \ 0 \\ 1 \ 0 \\
\end{array}\right](y'+\alpha+\frac{1}{2},z'+\beta+\frac{1}{2})\  
\vartheta\left[\begin{array}{cc} 1 \ 0 \\ 1 \ 0 \\
\end{array}\right](y',z')\Big)
\nonumber\\
&&\hskip -20mm -\Big(\vartheta\left[\begin{array}{cc} 0 \ 1 \\ 1 \ 0 \\
\end{array}\right](y+\alpha+\frac{1}{2},z+\beta+\frac{1}{2})\  
\vartheta\left[\begin{array}{cc} 0 \ 1 \\ 1 \ 0 \\
\end{array}\right](y,z)\  
\vartheta\left[\begin{array}{cc} 0 \ 1 \\ 1 \ 0 \\
\end{array}\right](y'+\alpha+\frac{1}{2},z'+\beta+\frac{1}{2})\  
\vartheta\left[\begin{array}{cc} 0 \ 1 \\ 1 \ 0 \\
\end{array}\right](y',z')\Big)
\nonumber\\
&&\hskip -20mm +\Big(\vartheta\left[\begin{array}{cc} 1 \ 1 \\ 1 \ 0 \\
\end{array}\right](y+\alpha+\frac{1}{2},z+\beta+\frac{1}{2})\ 
\vartheta\left[\begin{array}{cc} 1 \ 1 \\ 1 \ 0 \\
\end{array}\right](y,z)\ 
\vartheta\left[\begin{array}{cc} 1 \ 1 \\ 1 \ 0 \\
\end{array}\right](y'+\alpha+\frac{1}{2},z'+\beta+\frac{1}{2})\ 
\vartheta\left[\begin{array}{cc} 1 \ 1 \\ 1 \ 0 \\
\end{array}\right](y',z')\Big)  . 
\nonumber\\
&& \label{A-10}
\end{eqnarray}
where we use the relation that the left-hand
side of Eq.(\ref{A-9}) does not change under the rename of 
$\alpha \leftrightarrow \beta$, $y \leftrightarrow z$, $y' \leftrightarrow z'$, 
$\tau_1 \leftrightarrow \tau_2$, then the first column and the second column
exchanged expression of the theta function in the right-hand side Eq.(\ref{A-9})
is equal to the original expression.\\
\noindent \underline{Another Relation}\\
Next we make another relation.
Then we replace $y\rightarrow y+1/2$, $y' \rightarrow y' +1/2$ in
Eq.(\ref{A-4}), and we have
\begin{eqnarray}
&&\hskip -20mm  \vartheta\left[\begin{array}{cc} 0 \ 0 \\ 0 \ 0 \\ 
\end{array}\right](\alpha,\beta)\ 
\vartheta\left[\begin{array}{cc} 0 \ 0 \\ 0 \ 0 \\ 
\end{array}\right](0,\frac{1}{2})\ 
\vartheta\left[\begin{array}{cc} 0 \ 0 \\ 0 \ 0 \\ 
\end{array}\right](y+y'+\alpha,z+z'+\beta)\ 
\vartheta\left[\begin{array}{cc} 0 \ 0 \\ 0 \ 0 \\ 
\end{array}\right](y-y',z-z'+\frac{1}{2})
\nonumber\\
&&\hskip -20mm -\vartheta\left[\begin{array}{cc} 1 \ 0 \\ 0 \ 0 \\
\end{array}\right](\alpha,\beta)\ 
\vartheta\left[\begin{array}{cc} 1 \ 0 \\ 0 \ 0 \\
\end{array}\right](0,\frac{1}{2})\ 
\vartheta\left[\begin{array}{cc} 1 \ 0 \\ 0 \ 0 \\
\end{array}\right](y+y'+\alpha,z+z'+\beta)\ 
\vartheta\left[\begin{array}{cc} 1 \ 0 \\ 0 \ 0 \\
\end{array}\right](y-y',z-z'+\frac{1}{2})
\nonumber\\
&&\hskip -20mm =\vartheta\left[\begin{array}{cc} 0 \ 0 \\ 1 \ 0 \\ 
\end{array}\right](y+\alpha,z+\beta+\frac{1}{2})\ 
\vartheta\left[\begin{array}{cc} 0 \ 0 \\ 1 \ 0 \\ 
\end{array}\right](y,z)\ 
\vartheta\left[\begin{array}{cc} 0 \ 0 \\ 1 \ 0 \\ 
\end{array}\right](y'+\alpha,z'+\beta+\frac{1}{2})\ 
\vartheta\left[\begin{array}{cc} 0 \ 0 \\ 1 \ 0 \\ 
\end{array}\right](y',z')
\nonumber\\
&&\hskip -20mm -\vartheta\left[\begin{array}{cc} 0 \ 1 \\ 1 \ 0 \\
\end{array}\right](y+\alpha,z+\beta+\frac{1}{2})\ 
\vartheta\left[\begin{array}{cc} 0 \ 1 \\ 1 \ 0 \\
\end{array}\right](y,z)\  
\vartheta\left[\begin{array}{cc} 0 \ 1 \\ 1 \ 0 \\
\end{array}\right](y'+\alpha,z'+\beta+\frac{1}{2})\  
\vartheta\left[\begin{array}{cc} 0 \ 1 \\ 1 \ 0\\
\end{array}\right](y',z')
\nonumber\\
&&\hskip -20mm +\vartheta\left[\begin{array}{cc} 1 \ 0 \\ 1 \ 0 \\
\end{array}\right](y+\alpha,z+\beta+\frac{1}{2})\  
\vartheta\left[\begin{array}{cc} 1 \ 0 \\ 1 \ 0 \\
\end{array}\right](y,z)\  
\vartheta\left[\begin{array}{cc} 1 \ 0 \\ 1 \ 0 \\
\end{array}\right](y'+\alpha,z'+\beta+\frac{1}{2})\  
\vartheta\left[\begin{array}{cc} 1 \ 0 \\ 1 \ 0 \\
\end{array}\right](y',z')
\nonumber\\
&&\hskip -20mm -\vartheta\left[\begin{array}{cc} 1 \ 1 \\ 1 \ 0 \\
\end{array}\right](y+\alpha,z+\beta+\frac{1}{2})\ 
\vartheta\left[\begin{array}{cc} 1 \ 1 \\ 1 \ 0 \\
\end{array}\right](y,z)\ 
\vartheta\left[\begin{array}{cc} 1 \ 1 \\ 1 \ 0 \\
\end{array}\right](y'+\alpha,z'+\beta+\frac{1}{2})\ 
\vartheta\left[\begin{array}{cc} 1 \ 1 \\ 1 \ 0 \\
\end{array}\right](y',z')  . 
\nonumber\\
&&\label{A-11}
\end{eqnarray}
Adding Eq.(\ref{A-4}) and Eq.(\ref{A-11}), and further replace
$\alpha \rightarrow \alpha+1/2$, we have
\begin{eqnarray}
&&\hskip -20mm   2\vartheta\left[\begin{array}{cc} 0 \ 0 \\ 0 \ 0 \\ 
\end{array}\right](\alpha+\frac{1}{2},\beta)\ 
\vartheta\left[\begin{array}{cc} 0 \ 0 \\ 0 \ 0 \\ 
\end{array}\right](0,\frac{1}{2})\ 
\vartheta\left[\begin{array}{cc} 0 \ 0 \\ 0 \ 0 \\ 
\end{array}\right](y+y'+\alpha+\frac{1}{2},z+z'+\beta)\ 
\vartheta\left[\begin{array}{cc} 0 \ 0 \\ 0 \ 0 \\ 
\end{array}\right](y-y',z-z'+\frac{1}{2})
\nonumber\\
&&\hskip -20mm =\vartheta\left[\begin{array}{cc} 0 \ 0 \\ 0 \ 0 \\ 
\end{array}\right](y+\alpha+\frac{1}{2},z+\beta+\frac{1}{2})\ 
\vartheta\left[\begin{array}{cc} 0 \ 0 \\ 0 \ 0 \\ 
\end{array}\right](y,z)\ 
\vartheta\left[\begin{array}{cc} 0 \ 0 \\ 0 \ 0 \\ 
\end{array}\right](y'+\alpha+\frac{1}{2},z'+\beta+\frac{1}{2})\ 
\vartheta\left[\begin{array}{cc} 0 \ 0 \\ 0 \ 0 \\ 
\end{array}\right](y',z')
\nonumber\\
&&\hskip -20mm -\vartheta\left[\begin{array}{cc} 0 \ 1 \\ 0 \ 0 \\
\end{array}\right](y+\alpha+\frac{1}{2},z+\beta+\frac{1}{2})\ 
\vartheta\left[\begin{array}{cc} 0 \ 1 \\ 0 \ 0 \\
\end{array}\right](y,z)\  
\vartheta\left[\begin{array}{cc} 0 \ 1 \\ 0 \ 0 \\
\end{array}\right](y'+\alpha+\frac{1}{2},z'+\beta+\frac{1}{2})\  
\vartheta\left[\begin{array}{cc} 0 \ 1 \\ 0 \ 0 \\
\end{array}\right](y',z')
\nonumber\\
&&\hskip -20mm +\vartheta\left[\begin{array}{cc} 1 \ 0 \\ 0 \ 0 \\
\end{array}\right](y+\alpha+\frac{1}{2},z+\beta+\frac{1}{2})\  
\vartheta\left[\begin{array}{cc} 1 \ 0 \\ 0 \ 0 \\
\end{array}\right](y,z)\  
\vartheta\left[\begin{array}{cc} 1 \ 0 \\ 0 \ 0 \\
\end{array}\right](y'+\alpha+\frac{1}{2},z'+\beta+\frac{1}{2})\  
\vartheta\left[\begin{array}{cc} 1 \ 0 \\ 0 \ 0 \\
\end{array}\right](y',z')
\nonumber\\
&&\hskip -20mm -\vartheta\left[\begin{array}{cc} 1 \ 1 \\ 0 \ 0 \\
\end{array}\right](y+\alpha+\frac{1}{2},z+\beta+\frac{1}{2})\ 
\vartheta\left[\begin{array}{cc} 1 \ 1 \\ 0 \ 0 \\
\end{array}\right](y,z)\ 
\vartheta\left[\begin{array}{cc} 1 \ 1 \\ 0 \ 0 \\
\end{array}\right](y'+\alpha+\frac{1}{2},z'+\beta+\frac{1}{2})\ 
\vartheta\left[\begin{array}{cc} 1 \ 1 \\ 0 \ 0 \\
\end{array}\right](y',z')
\nonumber\\
&&\hskip -20mm +\vartheta\left[\begin{array}{cc} 0 \ 0 \\ 1 \ 0 \\ 
\end{array}\right](y+\alpha+\frac{1}{2},z+\beta+\frac{1}{2})\ 
\vartheta\left[\begin{array}{cc} 0 \ 0 \\ 1 \ 0 \\ 
\end{array}\right](y,z)\ 
\vartheta\left[\begin{array}{cc} 0 \ 0 \\ 1 \ 0 \\ 
\end{array}\right](y'+\alpha+\frac{1}{2},z'+\beta+\frac{1}{2})\ 
\vartheta\left[\begin{array}{cc} 0 \ 0 \\ 1 \ 0 \\ 
\end{array}\right](y',z')
\nonumber\\
&&\hskip -20mm -\vartheta\left[\begin{array}{cc} 0 \ 1 \\ 1 \ 0 \\
\end{array}\right](y+\alpha,+\frac{1}{2}z+\beta+\frac{1}{2})\ 
\vartheta\left[\begin{array}{cc} 0 \ 1 \\ 1 \ 0 \\
\end{array}\right](y,z)\  
\vartheta\left[\begin{array}{cc} 0 \ 1 \\ 1 \ 0 \\
\end{array}\right](y'+\alpha+\frac{1}{2},z'+\beta+\frac{1}{2})\  
\vartheta\left[\begin{array}{cc} 0 \ 1 \\ 1 \ 0\\
\end{array}\right](y',z')
\nonumber\\
&&\hskip -20mm +\vartheta\left[\begin{array}{cc} 1 \ 0 \\ 1 \ 0 \\
\end{array}\right](y+\alpha+\frac{1}{2},z+\beta+\frac{1}{2})\  
\vartheta\left[\begin{array}{cc} 1 \ 0 \\ 1 \ 0 \\
\end{array}\right](y,z)\  
\vartheta\left[\begin{array}{cc} 1 \ 0 \\ 1 \ 0 \\
\end{array}\right](y'+\alpha+\frac{1}{2},z'+\beta+\frac{1}{2})\  
\vartheta\left[\begin{array}{cc} 1 \ 0 \\ 1 \ 0 \\
\end{array}\right](y',z')
\nonumber\\
&&\hskip -20mm -\vartheta\left[\begin{array}{cc} 1 \ 1 \\ 1 \ 0 \\
\end{array}\right](y+\alpha+\frac{1}{2},z+\beta+\frac{1}{2})\ 
\vartheta\left[\begin{array}{cc} 1 \ 1 \\ 1 \ 0 \\
\end{array}\right](y,z)\ 
\vartheta\left[\begin{array}{cc} 1 \ 1 \\ 1 \ 0 \\
\end{array}\right](y'+\alpha+\frac{1}{2},z'+\beta+\frac{1}{2})\ 
\vartheta\left[\begin{array}{cc} 1 \ 1 \\ 1 \ 0 \\
\end{array}\right](y',z')  .  
\nonumber\\
&& \label{A-12}
\end{eqnarray}
Then we add Eq.(\ref{A-10}) and \Big( Eq.(\ref{A-12})\Big).
\begin{eqnarray}
&&\hskip -20mm   \Big(\vartheta\left[\begin{array}{cc} 0 \ 0 \\ 0 \ 0 \\ 
\end{array}\right](\alpha+\frac{1}{2},\beta)\ 
\vartheta\left[\begin{array}{cc} 0 \ 0 \\ 0 \ 0 \\ 
\end{array}\right](0,\frac{1}{2})\ 
\vartheta\left[\begin{array}{cc} 0 \ 0 \\ 0 \ 0 \\ 
\end{array}\right](y+y'+\alpha+\frac{1}{2},z+z'+\beta)\ 
\vartheta\left[\begin{array}{cc} 0 \ 0 \\ 0 \ 0 \\ 
\end{array}\right](y-y',z-z'+\frac{1}{2})\Big)
\nonumber\\
&&\hskip -20mm =\Big(\vartheta\left[\begin{array}{cc} 0 \ 0 \\ 0 \ 0 \\ 
\end{array}\right](y+\alpha+\frac{1}{2},z+\beta+\frac{1}{2})\ 
\vartheta\left[\begin{array}{cc} 0 \ 0 \\ 0 \ 0 \\ 
\end{array}\right](y,z)\ 
\vartheta\left[\begin{array}{cc} 0 \ 0 \\ 0 \ 0 \\ 
\end{array}\right](y'+\alpha+\frac{1}{2},z'+\beta+\frac{1}{2})\ 
\vartheta\left[\begin{array}{cc} 0 \ 0 \\ 0 \ 0 \\ 
\end{array}\right](y',z')\Big)
\nonumber\\
&&\hskip -20mm -\Big(\vartheta\left[\begin{array}{cc} 0 \ 1 \\ 0 \ 0 \\
\end{array}\right](y+\alpha+\frac{1}{2},z+\beta+\frac{1}{2})\ 
\vartheta\left[\begin{array}{cc} 0 \ 1 \\ 0 \ 0 \\
\end{array}\right](y,z)\  
\vartheta\left[\begin{array}{cc} 0 \ 1 \\ 0 \ 0 \\
\end{array}\right](y'+\alpha+\frac{1}{2},z'+\beta+\frac{1}{2})\  
\vartheta\left[\begin{array}{cc} 0 \ 1 \\ 0 \ 0 \\
\end{array}\right](y',z')\Big)
\nonumber\\
&&\hskip -20mm +\Big(\vartheta\left[\begin{array}{cc} 0 \ 0 \\ 1 \ 0 \\ 
\end{array}\right](y+\alpha+\frac{1}{2},z+\beta+\frac{1}{2})\ 
\vartheta\left[\begin{array}{cc} 0 \ 0 \\ 1 \ 0 \\ 
\end{array}\right](y,z)\ 
\vartheta\left[\begin{array}{cc} 0 \ 0 \\ 1 \ 0 \\ 
\end{array}\right](y'+\alpha+\frac{1}{2},z'+\beta+\frac{1}{2})\ 
\vartheta\left[\begin{array}{cc} 0 \ 0 \\ 1 \ 0 \\ 
\end{array}\right](y',z')\Big)
\nonumber\\
&&\hskip -20mm -\Big(\vartheta\left[\begin{array}{cc} 0 \ 1 \\ 1 \ 0 \\
\end{array}\right](y+\alpha+\frac{1}{2},z+\beta+\frac{1}{2})\ 
\vartheta\left[\begin{array}{cc} 0 \ 1 \\ 1 \ 0 \\
\end{array}\right](y,z)\  
\vartheta\left[\begin{array}{cc} 0 \ 1 \\ 1 \ 0 \\
\end{array}\right](y'+\alpha+\frac{1}{2},z'+\beta+\frac{1}{2})\  
\vartheta\left[\begin{array}{cc} 0 \ 1 \\ 1 \ 0\\
\end{array}\right](y',z')\Big) . 
\nonumber\\
&& \label{A-13}
\end{eqnarray}
By replacing $y \rightarrow y+1$, $z \rightarrow z+1/2$, 
$y' \rightarrow y'+1/2$, $z' \rightarrow z'+1/2$, we have
\begin{eqnarray}
&&\hskip -20mm   \Big(\vartheta\left[\begin{array}{cc} 0 \ 0 \\ 1 \ 0 \\ 
\end{array}\right](\alpha,\beta)\ 
\vartheta\left[\begin{array}{cc} 0 \ 0 \\ 0 \ 1 \\ 
\end{array}\right](0,0)\ 
\vartheta\left[\begin{array}{cc} 0 \ 0 \\ 0 \ 0 \\ 
\end{array}\right](y+y'+\alpha,z+z'+\beta)\ 
\vartheta\left[\begin{array}{cc} 0 \ 0 \\ 1 \ 1 \\ 
\end{array}\right](y-y',z-z')\Big)
\nonumber\\
&&\hskip -20mm =\Big(\vartheta\left[\begin{array}{cc} 0 \ 0 \\ 1 \ 0 \\ 
\end{array}\right](y+\alpha,z+\beta)\ 
\vartheta\left[\begin{array}{cc} 0 \ 0 \\ 0 \ 1 \\ 
\end{array}\right](y,z)\ 
\vartheta\left[\begin{array}{cc} 0 \ 0 \\ 0 \ 0 \\ 
\end{array}\right](y'+\alpha,z'+\beta)\ 
\vartheta\left[\begin{array}{cc} 0 \ 0 \\ 1 \ 1 \\ 
\end{array}\right](y',z')\Big)
\nonumber\\
&&\hskip -20mm -\Big(\vartheta\left[\begin{array}{cc} 0 \ 1 \\ 1 \ 0 \\
\end{array}\right](y+\alpha,z+\beta)\ 
\vartheta\left[\begin{array}{cc} 0 \ 1 \\ 0 \ 1 \\
\end{array}\right](y,z)\  
\vartheta\left[\begin{array}{cc} 0 \ 1 \\ 0 \ 0 \\
\end{array}\right](y'+\alpha,z'+\beta)\  
\vartheta\left[\begin{array}{cc} 0 \ 1 \\ 1 \ 1 \\
\end{array}\right](y',z')\Big)
\nonumber\\
&&\hskip -20mm +\Big(\vartheta\left[\begin{array}{cc} 0 \ 0 \\ 0 \ 0 \\ 
\end{array}\right](y+\alpha,z+\beta)\ 
\vartheta\left[\begin{array}{cc} 0 \ 0 \\ 1 \ 1 \\ 
\end{array}\right](y,z)\ 
\vartheta\left[\begin{array}{cc} 0 \ 0 \\ 1 \ 0 \\ 
\end{array}\right](y'+\alpha,z'+\beta)\ 
\vartheta\left[\begin{array}{cc} 0 \ 0 \\ 0 \ 1 \\ 
\end{array}\right](y',z')\Big)
\nonumber\\
&&\hskip -20mm -\Big(\vartheta\left[\begin{array}{cc} 0 \ 1 \\ 0 \ 0 \\
\end{array}\right](y+\alpha,z+\beta)\ 
\vartheta\left[\begin{array}{cc} 0 \ 1 \\ 1 \ 1 \\
\end{array}\right](y,z)\  
\vartheta\left[\begin{array}{cc} 0 \ 1 \\ 1 \ 0 \\
\end{array}\right](y'+\alpha,z'+\beta)\  
\vartheta\left[\begin{array}{cc} 0 \ 1 \\ 0 \ 1\\
\end{array}\right](y',z')\Big) . 
\label{A-14}
\end{eqnarray}
This is the Kossak's Fundamental Relation 2) of Eq.(\ref{e2-6}).\\
\noindent \underline{Kossak's Fundamental Relation 3):}\\
From the above, we also have the Kossak's Fundamental
relation 3) of Eq.(\ref{e2-7}) in the form
\begin{eqnarray}
&&\hskip -20mm   \Big(\vartheta\left[\begin{array}{cc} 0 \ 0 \\ 0 \ 1 \\ 
\end{array}\right](\alpha,\beta)\ 
\vartheta\left[\begin{array}{cc} 0 \ 0 \\ 1 \ 0 \\ 
\end{array}\right](0,0)\ 
\vartheta\left[\begin{array}{cc} 0 \ 0 \\ 0 \ 0 \\ 
\end{array}\right](y+y'+\alpha,z+z'+\beta)\ 
\vartheta\left[\begin{array}{cc} 0 \ 0 \\ 1 \ 1 \\ 
\end{array}\right](y-y',z-z')\Big)
\nonumber\\
&&\hskip -20mm =\Big(\vartheta\left[\begin{array}{cc} 0 \ 0 \\ 0 \ 1 \\ 
\end{array}\right](y+\alpha,z+\beta)\ 
\vartheta\left[\begin{array}{cc} 0 \ 0 \\ 1 \ 0 \\ 
\end{array}\right](y,z)\ 
\vartheta\left[\begin{array}{cc} 0 \ 0 \\ 0 \ 0 \\ 
\end{array}\right](y'+\alpha,z'+\beta)\ 
\vartheta\left[\begin{array}{cc} 0 \ 0 \\ 1 \ 1 \\ 
\end{array}\right](y',z')\Big)
\nonumber\\
&&\hskip -20mm -\Big(\vartheta\left[\begin{array}{cc} 1 \ 0 \\ 0 \ 1 \\
\end{array}\right](y+\alpha,z+\beta)\ 
\vartheta\left[\begin{array}{cc} 1 \ 0 \\ 1 \ 0 \\
\end{array}\right](y,z)\  
\vartheta\left[\begin{array}{cc} 1 \ 0 \\ 0 \ 0 \\
\end{array}\right](y'+\alpha,z'+\beta)\  
\vartheta\left[\begin{array}{cc} 1 \ 0 \\ 1 \ 1 \\
\end{array}\right](y',z')\Big)
\nonumber\\
&&\hskip -20mm +\Big(\vartheta\left[\begin{array}{cc} 0 \ 0 \\ 0 \ 0 \\ 
\end{array}\right](y+\alpha,z+\beta)\ 
\vartheta\left[\begin{array}{cc} 0 \ 0 \\ 1 \ 1 \\ 
\end{array}\right](y,z)\ 
\vartheta\left[\begin{array}{cc} 0 \ 0 \\ 0 \ 1 \\ 
\end{array}\right](y'+\alpha,z'+\beta)\ 
\vartheta\left[\begin{array}{cc} 0 \ 0 \\ 1 \ 0 \\ 
\end{array}\right](y',z')\Big)
\nonumber\\
&&\hskip -20mm -\Big(\vartheta\left[\begin{array}{cc} 1 \ 0 \\ 0 \ 0 \\
\end{array}\right](y+\alpha,z+\beta)\ 
\vartheta\left[\begin{array}{cc} 1 \ 0 \\ 1 \ 1 \\
\end{array}\right](y,z)\  
\vartheta\left[\begin{array}{cc} 1 \ 0 \\ 0 \ 1 \\
\end{array}\right](y'+\alpha,z'+\beta)\  
\vartheta\left[\begin{array}{cc} 1 \ 0 \\ 1 \ 0\\
\end{array}\right](y',z')\Big) , 
\label{A-15}
\end{eqnarray}
where we use the relation that the left-hand
side of Eq.(\ref{A-14}) does not change under the rename of 
$\alpha \leftrightarrow \beta$, $y \leftrightarrow z$, $y' \leftrightarrow z'$, 
$\tau_1 \leftrightarrow \tau_2$, then the first column and the second column
exchanged expression of the theta function in the right-hand side of Eq.(\ref{A-14})
is equal to the original expression.

\end{document}